\documentclass{LMCS}
\usepackage{hyperref}
\usepackage{graphicx}
\usepackage{amssymb}
\usepackage{amsmath}
\usepackage{stmaryrd}
\usepackage{epsfig}
\usepackage{color}
\usepackage{listings}
\let \lst \lstinline

\usepackage{xspace}
\usepackage{mathpartir,enumerate,hyperref}

\newcommand{\forget}[1]{}
\newcommand{\etal}{\emph{et al.}\@\xspace}
\newcommand{\pair}[2]{#1\mathord{\texttt{,}}#2}
\renewcommand{\exp}{e}
\newcommand{\resetequationcounter}{\setcounter{equation}{0}}
\newcommand{\const}{\kappa}
\newcommand{\pt}{\pi}
\newcommand{\ptbis}{\omega}
\newcommand{\ptsbis}{\Omega}
\newcommand{\pts}{\Pi}
\newcommand{\lessprecise}{\preceq}
\newcommand{\ins}[1]{{\sf Ins}(#1)}

\newcommand{\rln}[1]{\text{\small\sc #1}}
\newcommand{\ie}{\emph{i.e.}\@\xspace}
\newcommand{\eg}{\emph{e.g.}\@\xspace}
\newcommand{\jocaml}{\textrm{JoCaml}\xspace}
\newcommand{\ocaml}{\textrm{OCaml}\xspace}
\newcommand{\erlang}{\textrm{Erlang}\xspace}            
\newcommand{\csharp}{C$^{\sharp}$\xspace}
\newcommand{\haskell}{\textrm{Haskell}\xspace}
\newcommand{\comega}{C$\omega$\xspace}
\newcommand{\funnel}{\textrm{Funnel}\xspace}
\newcommand{\scala}{\textrm{Scala}\xspace}
\newcommand{\java}{\textrm{Java}\xspace}
\newcommand{\joinjava}{\textrm{JoinJava}\xspace}
\newcommand{\rcham}{\textrm{RCHAM}\xspace}
\newcommand{\cham}{\textrm{CHAM}\xspace}
\newcommand{\Id}{\emph{Id}\@\xspace}
\newcommand{\kwd}[1]{\ensuremath{\mathbf{#1}}}
\newcommand{\prefix}[1]{\mathopen{}\mathrel{\kwd {#1}}}
\newcommand{\infix}[1]{\mathrel{\kwd {#1}}}

\newcommand{\id}[1]{\textit{#1}}

\newcommand{\set}[1]{\{#1\}}
\newcommand{\lubop}{\mathop{\uparrow}}
\newcommand{\lub}[2]{#1 \lubop #2}
\newcommand{\cons}[2]{#1\mathord{\texttt{::}}#2}
\newcommand{\nil}{\texttt{[]}}
\newcommand{\single}[1]{\texttt{[}#1\texttt{]}}
\newcommand{\repr}[2]{#1 \mathop{\updownarrow} #2}
\renewcommand{\_}{\mathord{\rule[-.25ex]{1ex}{.15ex}}}
\newcommand{\appnv}[4]{{#1}(#2, #3, \ldots{} , #4)}
\newcommand{\V}{\mathcal{V}}
\newcommand{\E}{\mathcal{E}}

\newcommand{\nullp}{0}
\newcommand{\defineas}{\;\stackrel{\mathrm{def}}=\;}
\newcommand{\C}[1]{\llbracket#1\rrbracket}
\newcommand{\CPLambda}[1]{\llbracket#1\rrbracket_{\lambda}}
\newcommand{\para}[2]{#1\mathop{\&}#2}
\newcommand{\dis}[2]{#1\infix{or}#2}
\newcommand{\define}[2]{\prefix{def} #1 \infix{in} #2}
\newcommand{\matchfour}[4]{\prefix{match}  #1 \infix{with} #2_1
  \rightarrow #3_1 \mid \ldots \mid #2_{#4} \rightarrow #3_{#4}}
\newcommand{\matchthree}[3]{\prefix{match} #1 \infix{with} \mid^{i
    \in I} #2_i \rightarrow #3_i}
\newcommand{\matchone}[3]{\prefix{match} #1 \infix{with} #2
  \rightarrow #3} 
\newcommand{\matchC}[3]{\prefix{match} #1 \infix{with} \mid^{i
    \in I} #2_i \rightarrow \C{#3_i}}
\newcommand{\reaction}[2]{#1 \triangleright #2}
\newcommand{\soup}[2]{#1\vdash#2}
\newcommand{\heat}{\rightharpoonup}
\newcommand{\cool}{\rightharpoondown}
\newcommand{\reduces}{\longrightarrow}
\newcommand{\hc}{\rightleftharpoons}
\newcommand{\dd}{\mathcal{D}}
\newcommand{\pp}{\mathcal{P}}
\newcommand{\wide}[1]{\mathrel {{}#1{}}}
\newcommand{\hcred}{\overset{\heat}{\underset{\cool}{\reduces}}}
\newcommand{\sub}[2]{#1#2}
\newcommand{\restrict}{\mathbin{\upharpoonright}}
\newcommand{\composite}[2]{#1 \mathop{\circ} #2} 
\newcommand{\wbc}{\mathrel{\approx}} 
\newcommand{\R}{\mathrel{\mathcal{R}}}
\newcommand{\pseq}{\;\mathop{\Bumpeq}\;}
\newcommand{\weakbarb}[2]{#1\!\Downarrow_{#2}}
\newcommand{\barb}[2]{#1\!\downarrow_{#2}}
\newcommand{\econtext}[1]{E[#1]}
\newcommand{\mctxt}[1]{\prefix{match} \exp \infix{with} \ldots \mid \pi_k
  \rightarrow #1 \mid \ldots}
\newcommand{\jctxt}[2][R]{\define{\dis{\reaction{J}{#2}}{D}}{#1}}
\newcommand{\preds}[1]{I(#1)}
\newcommand{\fv}[1]{\mbox{\sf fv}[#1]}
\newcommand{\dv}[1]{\mbox{\sf dv}[#1]}
\newcommand{\rv}[1]{\mbox{\sf rv}[#1]}
\def \dom #1{\mbox{\sf dom}(#1)}
\newcommand{\bisi}[1]{\widehat{#1}}

\def\doi{4 (1:7) 2008}
\lmcsheading%
{\doi}
{1--41}
{}
{}
{Jan.~25, 2007}
{Mar.~21, 2008}
{}

\begin{document}

\title{Algebraic Pattern Matching in Join Calculus\rsuper *}

\author[Q.~Ma]{Qin Ma\rsuper a}
\address{{\lsuper a}OFFIS, Escherweg 2, 26121 Oldenburg, Germany}
\email{Qin.Ma@offis.de}

\author[L.~Maranget]{Luc Maranget\rsuper b}
\address{{\lsuper b}INRIA-Rocquencourt, BP 105, 78153 Le Chesnay Cedex, France}
\email{Luc.Maranget@inria.fr}

\titlecomment{{\lsuper *}Extended version of~\cite{MaMaranget2004pattern}}

\begin{abstract}
  We propose an extension of the join calculus with pattern matching on
  algebraic data types. Our initial motivation is twofold: to provide an
  intuitive semantics of the interaction between concurrency and pattern
  matching; to define a practical compilation scheme from extended join
  definitions into ordinary ones plus ML pattern matching. To assess the
  correctness of our compilation scheme, we develop a theory of the applied
  join calculus, a calculus with value passing and value matching. We
  implement this calculus as an extension of the current \jocaml system.
\end{abstract}

\maketitle


\section{Introduction}

\label{sec.intro}

The join calculus~\cite{Fournet98:PhD,FournetGonthier96} is a process
calculus in the tradition of the $\pi$-calculus of Milner
\etal~\cite{MPW92}. One distinctive feature of join calculus is the
simultaneous definition of all receptors on several channels through
\emph{join definitions}. A~join definition is structured as a list of
\emph{reaction rules}, with each reaction rule being a pair of one
\emph{join pattern} and one \emph{guarded process}. A join pattern is
in turn a list of channel names (with formal arguments), specifying
the synchronization among those channels: namely, a join pattern is
matched only if there are messages present on all its channels.
Finally, the reaction rules of one join definition define competing
behaviors with a non-deterministic choice of which guarded process to
trigger when several join patterns are satisfied.

In this paper, we extend the matching mechanism of join patterns, such
that \emph{message} contents are also taken into account. As an
example, let us consider the following list-based implementation of a
concurrent stack:\footnote{We use the \ocaml syntax for lists,
  with \emph{Nil} being $\nil$ and \emph{Cons} being the infix
  $\cons{}{}$.}
\begin{lstlisting}{Join}
def pop(r) & State(x::xs) |> r(x) & State(xs)
 or push(v) & State(ls) |> State (v::ls) 
 in State([]) & $\ldots$
\end{lstlisting}
The second join pattern \lst"push(v) & State(ls)" is an
\emph{ordinary} one: it is matched whenever there are messages on both
\lst"State" and \lst"push". By contrast, the first join pattern is an
\emph{extended} one, where the formal argument of channel \lst"State"
is an \emph{algebraic pattern}, matched only by messages that are
cons cells.  Thus, when the stack is empty (\ie, when message
$\nil$ is pending on channel \lst"State"), pop requests are delayed.
Note that we follow the convention that capitalized channels are
private: only \lst"push" and \lst"pop" will be visible outside.

A similar stack can be implemented without using extended join
patterns, but instead, using an extra private channel and ML pattern
matching in guarded processes:
\begin{lstlisting}{Join}
def pop(r) & Some(ls)  |> match ls with 
                           | [x] -> r(x) & Empty() 
                           | y::x::xs -> r(y) & Some(x::xs)
 or push(v) & Empty() |> Some ([v])
 or push(v) & Some(ls) |> Some (v::ls)
 in Empty() & $\ldots$
\end{lstlisting}
This second definition encodes the empty/non-empty status of
the stack as a message on channels \lst|Empty| and \lst|Some| respectively.
Pop requests on an empty stack are still delayed, since there is
no rule for the join pattern \lst|pop(r) & Empty()|.
The second definition obviously requires more programming
effort. Moreover, it is not immediately apparent that messages on
\lst"Some" are non-empty lists, and that the partial ML pattern
matching thus never fails.

Join definitions with (constant) pattern arguments appear informally
in functional nets~\cite{odersky:esop2000}. Here we generalize this
idea to full algebraic patterns. A similar attempt has also been
scheduled by Benton \etal as an interesting future work for
\comega~\cite{Cw}.

The new semantics is a smooth extension, since both join pattern
matching and pattern matching rest upon classical substitution (or
semi-unification).  However, an efficient implementation is more
involved. Our idea is to address this issue by transforming programs
whose definitions contain extended join patterns into equivalent
programs whose definitions use ordinary join patterns and whose
guarded processes use ML pattern matching. Doing so, we leave most of
the burden of pattern matching compilation to an ordinary ML~pattern
matching compiler. However, such a transformation is far from obvious.
More specifically, there is a gap between (extended) join pattern
matching, which is non-deterministic, and ML pattern matching, which
is deterministic (following the ``first match policy''). For example,
in our definition of a concurrent stack with extended join patterns,
\lst"State(ls)" is still matched by any message on \lst"State",
regardless of the presence of the more precise \lst"State(x::xs)" in
the competing reaction rule that precedes it.  Our solution to this
problem relies on partitioning matching values into non-intersecting
sets.  In the case of our concurrent stack, those sets simply are the
singleton $\{\lst"[]"\}$ and the set of non-empty lists.  Then,
pattern \lst"State(ls)" is matched by values from both sets, while
pattern \lst"State(x::xs)" is matched only by values of the second
set.

The rest of the paper is organized as follows: Section~\ref{sec.pt}
first gives a brief review of algebraic patterns and ML pattern
matching. Section~\ref{sec.joinpi} presents the applied join calculus
--- an extension of join with algebraic pattern matching.
We introduce the semantics and the appropriate equivalence relations.
Section~\ref{sec.trans-idea} informally explains the key ideas to
transform the extension to the ordinary join calculus, and especially how
we deal with the nondeterminism problem.
Section~\ref{sec.trans} formalizes the transformation as a
compilation scheme and presents the algorithm which essentially works
by building a meet semi-lattice of patterns.  We go through a complete
example in Section~\ref{sec.example}, and finally, we deal with the
correctness of the compilation scheme in Section~\ref{sec.correct}.
Implementation has been carried out as an extension of the \jocaml system.
We discuss the issues that have arisen during the
implementation work in Section~\ref{sec.imp}.

An earlier version of this paper (lacking the detailed proofs and the
discussion of the implementation) appeared
as~\cite{MaMaranget2004pattern}.

\section{Algebraic data types and ML pattern matching}
\label{sec.pt}

This section serves as a brief introduction to algebraic data types
and ML pattern matching. Interested readers are referred to
\cite{warning,LefessantMarangetPattern} for further details.

\subsection{Algebraic data types}
\label{subsec.pt}

In functional languages, new types can be introduced by using
\emph{data type definitions} and such types are algebraic
data types.
For example, using \ocaml syntax, binary trees can be defined as follows:
\begin{lstlisting}{Join}
type tree = Empty | Leaf of int | Node of tree * tree
\end{lstlisting}
The \emph{complete signature} of type \lst"tree" has three
\emph{constructors}: \lst"Empty", \lst"Leaf", and \lst"Node", which
are used to build the values of this type. Every constructor has an
arity, \ie the number of arguments it requires and meanwhile specifies
the corresponding types of each argument. In this definition,
\lst"Empty" is of arity zero, \lst"Leaf" is of arity one (and accepts
integer arguments),
and \lst"Node" is of arity two (both its arguments being themselves of
type \lst"tree"). A constructor of zero arity is sometimes called a
\emph{constant constructor}.

Most native ML data types can be seen as particular instances of
algebraic data types. For example, lists are defined by two
constructors: constant \lst"Nil" (written $\nil$) for empty
lists and \lst"Cons" (written as the infix $\cons{}{}$) for
nonempty ones; pairs are defined by one constructor with arity two,
(written as the infix ``$\pair{}{}$''); and integers
are defined by infinitely many (or $2^{31}$) constant constructors.

Formally, the algebraic values (for short values) of type~$t$ are well-typed terms built from the constructors of $t$.
``Well-typed'' here means correct with respect to constructor arity
and argument types.
Assuming a countable set of identifiers for constructors,
ranged over by $\const$, we give the formal definition of values as
follows:
$$
\begin{array}{rcll}
  v &::=& &  \textbf{Algebraic values}\\
  && \appnv{\const}{v_1}{v_2}{v_n} & \ \textrm{$\const$ of arity $n\geq 0$} 
\end{array}
$$
Type correctness is left implicit: we shall consider well typed terms
only.

Algebraic patterns (for short patterns) of type $t$ are also well-typed
terms built from the constructors of $t$, but with
variables.\footnote{We freely replace variables
whose names are of no importance by wildcards~``$\_$''.} The formal
definition of patterns is given as follows.$$
\begin{array}{rcll}
  \pt &::=& &  \textbf{Algebraic patterns}\\
  && x & \ \textrm{variable} \\ 
  && \appnv{\const}{\pt_1}{\pt_2}{\pt_n} & \ \textrm{$\const$ of arity $n\geq 0$} 
\end{array}
$$
We further require all variables in
a pattern to be pairwise distinct, that is, we only consider
\emph{linear} patterns.

Again, we assume a typed context. More precisely, we rely on the ML type
system to guarantee that values and patterns are well-typed.
Moreover, we rely on a ML type inferer to enrich syntax with
explicit types (which we leave implicit), and consider that the
type of any syntactic structure is available whenever needed.
Doing so, we focus on our main issue and avoid complications that would
be of little explanatory value.

Patterns are used to discriminate values according to their
structures. More specifically, a pattern denotes a set of values that
have a common prefix specified by the pattern. We say a value~$v$ (of
type~$t$) is an \emph{instance} of pattern~$\pt$ (of type~$t$), or
that $v$ matches $\pt$, when $\pt$ describes the prefix of $v$, in other words,
when there exists a substitution $\sigma$, such that ${\pi}{\sigma} =
v$.  For linear patterns, the instance relation can be defined
inductively as follows:
\begin{defi}[Instance]
  Let $\pt$ be a pattern and $v$ be a value, such that $\pt$ and $v$
  have the same type, the instance relation $\pt\lessprecise v$ is
  defined as:
  $$
  \begin{array}{rcll}
    \_ & \lessprecise & v & \\
    \const(\pt_1,\ldots,\pt_n)  & \lessprecise & \const(v_1,\ldots,v_n) & 
    \mbox{iff $\pt_i\lessprecise v_i$ for all $1 \leq i\leq n$}
  \end{array}
  $$
\end{defi}
\noindent We write $\ins{\pt}$ for the set of the instances of
pattern~$\pt$. The instance relation induces the following relations
among patterns. These relations apply to patterns $\pt_1$ and $\pt_2$
that have the same type.
\begin{defi}[Pattern relations] \hfill
\begin{enumerate}[$\bullet$]
\item Patterns $\pt_1$ and $\pt_2$ are {\bf compatible} when they
  share at least one instance. Otherwise $\pt_1$ and $\pt_2$ are {\bf
    incompatible} written $\pt_1 \# \pt_2$. Two compatible patterns
  admit a least upper bound written $\lub{\pt_1}{\pt_2}$, whose
  instance set is $\ins{\pt_1} \cap \ins{\pt_2}$.
\item Pattern $\pt_1$ is {\bf less precise} than pattern $\pt_2$
  written $\pt_1 \lessprecise \pt_2$ when $\ins{\pt_2} \subseteq
  \ins{\pt_1}$.
\item Patterns $\pt_1$ and $\pt_2$ are {\bf equivalent} written $\pt_1
  \equiv \pt_2$ when $\ins{\pt_1} = \ins{\pt_2}$.  If so, their least
  upper bound is their representative, written~$\repr{\pt_i}{\pt_2}$.
\end{enumerate}
\end{defi}
\noindent Note that we use the same notation $\lessprecise$ for both
relations: ``being an instance of'' (which is between a
pattern and a value) and ``being less precise'' (which is between two
patterns).
Indeed, values are in fact a special case of
patterns (with no variables), and in that case, both relations collapse.

The least upper bound of two patterns can be computed at the same time
when compatibility is checked by the following rules: $$
\begin{array}{rcllrcl}
\lub{\_}{\pt} & = & \pt \\
\lub{\pt}{\_} & = & \pt \\
\lub{\const(\pt_1,\ldots,\pt_n)}{\const(\ptbis_1,\ldots,\ptbis_n)} & = &
\const(\lub{\pt_1}{\ptbis_1}, \ldots, \lub{\pt_n}{\ptbis_n})
\end{array}
$$

Deciding the relation ``being less precise'' is more involved. Because
of typing, there exists nontrivial such relations, for instance
$(\pair{\_}{\_}) \lessprecise \_$.  The \jocaml compiler relies on an
efficient algorithm for this task, called the $\mathcal{U}$ algorithm,
with $\mathcal{U}$ standing for ``Usefulness''~\cite{warning}.
Algorithm~$\mathcal{U}$ takes two parameters: a list of patterns
$\pts$ and a pattern $\pt$, and returns a boolean. Roughly speaking, it
checks the usefulness of $\pt$ with respect to $\pts$. More
specifically, algorithm~$\mathcal{U}$ tests the existence of at least
one value $v$ such that $\pt$ admits~$v$ as an instance, and none of
the patterns in $\pts$ does.

From the point of view of algorithm~$\mathcal{U}$, deciding
the relation $\pt_1\lessprecise\pt_2$ amounts
to compute the \emph{negation} of $\mathcal{U}([\pt_1],\pt_2)$.
Namely, $\pt_1$ is less precise then $\pt_2$, if and only if
all the instances of~$\pt_2$ are instances of~$\pt_1$.
$$\pt_1\lessprecise\pt_2 \iff \mathcal{U}([\pt_1],\pt_2) = \kwd{useless}$$
We now give a simplified definition of algorithm~$\mathcal{U}$.
The simplified definition suffices for our needs and also conveys the
basic idea behind the algorithm.

Consider $\mathcal{U}([\pt_1],\pt_2)$, where
$\pt_1$ and~$\pt_2$ are patterns of a common type~$t$.
The following two cases are distinguished.\medskip

\noindent{\bf Case} $\pt_2 = \const(\ptbis_1,\ldots,\ptbis_n)$  \hfill
  \begin{enumerate}[$\bullet$]
  \item If $\pt_1 = \const(\gamma_1,\ldots,\gamma_n)$, then check if
    $\exists i, 1\leq i\leq n$, s.t.
    $\mathcal{U}([\gamma_i],\ptbis_i)$.
  \item If $\pt_1 = \const'(\gamma_1,\ldots,\gamma_n)$ and $\const
    \not= \const'$, then $\kwd{useful}$ (\ie $\kwd{false}$ for
    $\pt_1\lessprecise\pt_2$).
  \item If $\pt_1 = \_$, then $\kwd{useless}$ (\ie $\kwd{true}$ for
    $\pt_1\lessprecise\pt_2$).
  \end{enumerate}\medskip

\noindent{\bf Case} $\pt_2 = \_$  \hfill
  \begin{enumerate}[$\bullet$]
  \item If $\pt_1 = \_$, then $\kwd{useless}$ (\ie $\kwd{true}$ for
    $\pt_1\lessprecise\pt_2$).
  \item If $\pt_1 = \const(\gamma_1,\ldots,\gamma_n)$,
    \begin{enumerate}[$-$]
    \item if $\const$ is the unique constructor of type $t$, then
      check if $\exists i,1\leq i\leq n$, s.t.
      $\mathcal{U}([\gamma_i],\_)$.
    \item otherwise $\kwd{useful}$ (\ie $\kwd{false}$ for
      $\pt_1\lessprecise\pt_2$).
    \end{enumerate}
  \end{enumerate}\medskip

\noindent Once we can decide relation ``$\preceq$'', we can easily
decide pattern equivalence, since, by definition, $\pt_1\equiv\pt_2$
means $\pt_1\lessprecise\pt_2$ and $\pt_2\lessprecise\pt_1$.

\subsection{ML pattern matching}
\label{subsec.MLpt}

In ML, operating on algebraic data types is performed by the use of
the following \emph{\kwd{match} construct}
that we extend to processes ($Q_1$, $Q_2$ etc. below are processes of
the join calculus).
\begin{lstlisting}{Join}
  match $v$ with $\pt_1$ -> $Q_1$ | $\pt_2$ -> $Q_2$ | $\ldots$ | $\pt_n$ -> $Q_n$
\end{lstlisting}
Above, we attempt a matching of value~$v$ against a sequence of patterns
$\pt_1,\ldots,\pt_n$ of the same type. 

ML pattern matching is deterministic. It follows the ``first match
policy''. That is, when value~$v$ is an instance of more than one of
the patterns~$\pt_i$, the \kwd{match} construct chooses the one with
the smallest index~$i$.  This can be seen as checking patterns
$\pt_1$, $\pt_2$, \ldots , $\pt_n$ for admitting~$v$ as an instance
sequentially, stopping as soon as a match is found.  As a consequence,
pattern $\pt_i$ is matched only by the values in set
$\ins{\pt_i}\setminus(\textstyle{\bigcup_{1 \le j < i}}\ins{\pt_j})$.
Moreover, patterns in ML pattern matching also act as binding
constructs.  Once a successful match is found, say $\pt_k\lessprecise v$,
the variables in $\pt_k$ are all bound to the corresponding subterms
of $v$ in the guarded process~$Q_k$.

Additionally, we say a \kwd{match} construct is \emph{exhaustive}
when $\cup_{1 \le i \le n}
\ins{\pt_j}$ is the whole set of values of the considered type.
We accept non-exhaustive \kwd{match} constructs.

\section{The applied join calculus}
\label{sec.joinpi}

We define the applied join calculus by analogy with the applied
$\pi$-calculus~\cite{AppliedPi}. The applied join calculus inherits
its capabilities of communication and concurrency from pure join.
Moreover it supports algebraic value passing and algebraic pattern
matching in both join patterns and processes.

\subsection{Syntax and scopes}
\label{subsec.syntax}

The syntax of the applied join calculus is given in
Figure~\ref{fig.syntax}. As it is customary in process calculi
definitions, we assume an infinite set of
identifiers for variables, ranged over by $x, y, z$.

\begin{figure}[ht]
\centering
$$
\begin{array}{lcll}
P & ::=  &  & \textbf{Processes}\\
&& \nullp &  \ \textrm{null process}\\
&& x(\exp)& \ \textrm{message sending}\\
&& \para{P}{P} & \ \textrm{parallel}\\
&& \define{D}{P} & \ \textrm{definition}\\
&& \matchfour{\exp}{\pt}{P}{m} & \ \textrm{pattern matching}
\\
D & ::= & & \textbf{Join definitions}\\
&& \top & \ \textrm{empty definition}\\
&& \reaction{J}{P} & \ \textrm{reaction}\\
&& \dis{D}{D} & \ \textrm{disjunction}
\\
J & ::= & & \textbf{Join patterns}\\
&& x(\pt) & \ \textrm{message pattern}\\
&& \para{J}{J} & \ \textrm{synchronization}
\\
\pt & ::= & & \textbf{Algebraic patterns}\\
&& x & \ \textrm{variable}\\
&& \appnv{\const}{\pt_1}{\pt_2}{\pt_n} & \ \textrm{constructor pattern}
\\
\exp & ::= && \textbf{Expressions}\\
&& x & \ \textrm{variable}\\
&& \appnv{\const}{\exp_1}{\exp_2}{\exp_n} & \ \textrm{constructor expression}
\end{array}
$$
\caption{Syntax of the applied join calculus}\label{fig.syntax}
\end{figure}

With respect to pure join calculus, two new syntactic categories are
introduced: expressions and patterns.  At first glance, both
expressions~$e$ and patterns~$\pt$ are terms constructed from
variables and constructors, where $n$ stands for the arity of
constructor $\const$.  We make them different syntactic categories
for clarity, and also because we require patterns to be linear. We
also formalize the ML pattern matching in processes, as the new
\kwd{match} construct.  Moreover, in contrast to ordinary name passing
join calculus, there are two other, more radical, extensions: first,
in message sending, message contents become expressions as $x(\exp)$,
that is, we have value passing; second, when a channel name is defined
in a join pattern, in addition to the synchronization requirement, we
also specify what pattern the message content should satisfy by
$x(\pt)$.

There are two kinds of bindings: the definition process
$\define{D}{P}$ binds all the channel names defined in $D$ (written
$\dv{D}$) with scope $P$; and the reaction rule $\reaction{J}{P}$ or
the ML pattern matching $\matchfour{e}{\pi}{P}{m}$ bind all the local
variables (written $\rv{J}$ or $\rv{\pi_i}$) with scope $P$ or $P_i$,
$i \in \set{1, \ldots, m}$. The definition of the sets of defined
channel names $\dv{\cdot}$ is the same as in pure join.  By contrast,
the definition of sets $\rv{\cdot}$ has to be extended, so as to take
pattern arguments into account. Meanwhile, the definition of sets
$\fv{\cdot}$ should also be extended, to cater for the new \kwd{match}
process and expressions. We present the formal definitions of
$\dv{\cdot}$, $\rv{\cdot}$, and $\fv{\cdot}$ in
Figure~\ref{fig.scope}. In these rules, $\uplus$ is the disjoint
union, which expresses linearity constraints on both algebraic and
join patterns.

\begin{figure}[!p]
\centering
$$
\begin{array}{lrcl}
\multicolumn{4}{l}{\textbf{For algebraic patterns}:}\\
&\rv{x} & \defineas & \set{x} \\
&\rv{\appnv{\const}{\pi_1}{\pi_2}{\pi_n}} & \defineas 
&\rv{\pi_1} \uplus \rv{\pi_2} \uplus \ldots \uplus \rv{\pi_n}\\
\\
\multicolumn{4}{l}{\textbf{For expressions}:}\\
&\fv{x} & \defineas & \set{x} \\
&\fv{\appnv{\const}{\exp_1}{\exp_2}{\exp_n}} & \defineas 
&\fv{\exp_1} \cup \fv{\exp_2} \cup \ldots \cup \fv{\exp_n}\\
\\
\multicolumn{4}{l}{\textbf{For join patterns}:}\\
&\rv{x(\pi)} & \defineas & \rv{\pi}\\
&\rv{\para{J_1}{J_2}} & \defineas & \rv{J_1} \uplus \rv{J_2}\\
\\
&\dv{x(\pi)} & \defineas & \set{x}\\
&\dv{\para{J_1}{J_2}} & \defineas & \dv{J_1} \uplus \dv{J_2}\\
\\
\multicolumn{4}{l}{\textbf{For join definitions}:}\\
&\dv{\top} & \defineas & \emptyset\\
&\dv{\reaction{J}{P}} & \defineas & \dv{J}\\
&\dv{\dis{D_1}{D_2}} & \defineas & \dv{D_1} \cup \dv{D_2}\\
\\
&\fv{\top} & \defineas & \emptyset\\
&\fv{\reaction{J}{P}} & \defineas & \dv{J} \cup (\fv{P} \setminus \rv{J})\\
&\fv{\dis{D_1}{D_2}} & \defineas & \fv{D_1} \cup \fv{D_2}\\
\\
\multicolumn{4}{l}{\textbf{For processes}:}\\
&\fv{\nullp} & \defineas & \emptyset\\
&\fv{x(\exp)} & \defineas & \set{x} \cup \fv{\exp}\\
&\fv{\para{P_1}{P_2}} & \defineas & \fv{P_1} \cup \fv{P_2}\\
&\fv{\define{D}{P}} & \defineas & (\fv{D}\cup\fv{P}) \setminus \dv{D}\\
&\fv{\matchthree{e}{\pi}{P}} & \defineas & 
\fv{e} \cup (\bigcup_{i \in I} \fv{P_i} \setminus \rv{\pi_i}) \\
\\
\multicolumn{4}{l}{\textbf{For solutions}:}\\
&\dv{\dd} & \defineas & {\bigcup_{D \in \dd}\dv{D}}\\
&\fv{\dd} & \defineas & {\bigcup_{D \in \dd}\fv{D}} \\
&\fv{\pp} & \defineas & {\bigcup_{P \in \pp}\fv{P}}
\end{array}
$$
\caption{Bindings and scopes in the applied join
  calculus}\label{fig.scope}
\end{figure}

In applied join, values become of two kinds: channel names or
algebraic values.  We assume a type discipline in the style of the
type system of the
join-calculus~\cite{Fournet-Laneve-Maranget-Remy:typing-join},
extended with algebraic data types and the rule for ML pattern
matching.  Without making the type discipline more explicit, we
consider only well-typed terms (whose type we know), and assume that
substitutions preserve types. It should be observed that tuples are
now represented as a kind of constructed expressions and the arity
checking of polyadic join calculus is now replaced by a well-typing
assumption in applied join, which is thus monadic. One important
consequence of typing is that any (free) variable in a term possesses
a type and that we know this type.  Hence, we can discriminate between
those variables that are of a type of constructed values and those
that are of channel type.  Following the semantics of name passing
calculi such as join, we treat the latter kind of variables as channel
names, that is, values.  While, in any reasonable semantics, the
former kind of variables cannot be treated so.  We call a term
\emph{variable-closed} (\emph{closed} for short) when its free
variables are all of channel type, and otherwise \emph{open}.

\subsection{Chemical semantics}
\label{subsec.cham}

We establish the semantics following the reflexive chemical abstract
machine (\rcham) style --- the reflexive variant of
\cham~\cite{CHAM92}, whose states are chemical solutions. A
\emph{chemical solution} is a pair $\soup{\dd}{\pp}$,
where $\dd$ is a multiset of (active) join definitions, and
$\pp$ is a multiset of (running) processes.  Extending the
notion of closeness to solutions in the member-wise manner, we say a
solution is closed when all its active join definitions and running
processes are closed, namely, free variables are all of channel type.
We define semantics only on closed solutions. The chemical rewriting
rules are given in Figure~\ref{fig.cham},
consisting of two kinds as
in join: structural rules $\heat$ or $\cool$ represent the syntactical
rearrangement of the terms, and reduction rules $\reduces$ represent
the computation steps. We follow the convention to omit the part of
the solution that remains unchanged during rewrite.
This can also be expressed by the following context rule: $$
\infer[Context]
{\soup{\dd_0}{\pp_0} \wide\hcred \soup{\dd_1}{\pp_1}} {\soup{\dd,
    \dd_0}{\pp_0, \pp} \wide\hcred \soup{\dd, \dd_1}{\pp_1, \pp}} $$
where
$\hcred$ stands for either $\hc$ or $\reduces$, and
$\dd$ and $\pp$ are the independent context of the considered
subsolution. Rule~\rln{Str-Def} is a bit of exception because its side
condition actually requires the following relationship hold between
the rewriting part and its context: $\dv{D} \cap
(\fv{\dd}\cup\fv{\pp}) = \emptyset$.

Finally, it is perhaps to be noticed that, amongst the various,
slightly different, semantics of join-machines, we extend the one
of~\cite{Fournet-Laneve-Maranget-Remy:typing-join}, which is adapted
to static typing.  This means that we need to state explicitly that
$\kwd{or}$ is an associative-commutative operator.  As a consequence,
the notation $\dis{\reaction{J}{P}}{D}$ in rule~\rln{React} stands for
a definition that possesses a reaction rule whose pattern is~$J$.
\begin{figure}
\centering
$$
\begin{array}{lcrcl}
  \rln{Str-Null} && \soup{}{\nullp} & \hc & \soup{}{} \\
  \rln{Str-Par} && \soup{}{\para{P_1}{P_2}} & \hc & 
  \soup{}{P_1,P_2} \\
  \rln{Str-Top} && \soup{\top}{} & \hc & \soup{}{}\\
  \rln{Str-Def} && \soup{}{\define{D}{P}} & \hc & \soup{D}{P} \\
  \rln{React} && \soup{\dis{\reaction{J}{P}}{D}}{J\sigma}
  & \reduces & \soup{\dis{\reaction{J}{P}}{D}}{P\sigma}\\
  \rln{Match} &
  \multicolumn{4}{l}{\soup{}{\matchfour{\pt_i\eta}{\pt}{P}{m}}} \\
  &&& \reduces & \soup{}{P_i\eta}\\\\
  \mbox{Side conditions:}\\
  \ \rln{Str-Def}& \multicolumn{4}{l}{\mbox{$\dv{D}$ is fresh}}\\
  \ \rln{React}& 
  \multicolumn{4}{l}{\mbox{$\sigma$ substitutes closed expressions for $\rv{J}$}}\\
  \ \rln{Match}& 
  \multicolumn{4}{l}
  {\mbox{$\eta$ substitutes closed expressions for $\rv{\pt_i}$}}\\
  & \multicolumn{4}{l}{\mbox{and $\forall
      j < i, \pt_j \npreceq \pt_i\eta$}}\\
\end{array}
$$
\caption{RCHAM of the applied join calculus}\label{fig.cham}
\end{figure}

Matching of message contents against formal pattern arguments
is integrated in the substitution~$\sigma$ in rule \rln{React}.
As a consequence  this rule does not formally change
with respect to ordinary join calculus. However its semantical power has
much increased.
The \rln{Match} rule is new and expresses ML pattern matching.  
Its side condition enforces the first match policy.

According to the convention of processes as solutions, namely $P$ as
$\soup{}{P}$, the semantics is also defined on closed processes
in the following sense.
\begin{defi}
  Let $\hc^*$ denote the transitive closure of
  $\rightharpoonup\cup\rightharpoondown$, 
  \begin{enumerate}[(1)]
  \item $P \equiv Q \mbox{ iff } \soup{}{P} \hc^* \;\soup{}{Q}$
  \item $P \reduces Q \mbox{ iff } \soup{}{P} \hc^*\reduces\hc^*
    \;\soup{}{Q}$
  \end{enumerate}
\end{defi}
Subsequently, we have the following structural rule:
\begin{lem}\label{struct-rule}
  If $P\reduces Q$, $P\equiv P'$, and $Q\equiv Q'$, then $P'\reduces
  Q'$.
\end{lem}
\begin{proof}
  Trivially follow the definitions of $\equiv$ and $\reduces$, and the
  transitivity of $\hc^*$. \forget{\qed}
\end{proof}
\subsection{Equivalence relation}
\label{subsec.eq}

In this section, we equip the applied join calculus with equivalence
relations to allow reasoning over processes. The classical
notion of \emph{barbed congruence} is a sensible behavioral
equivalence based on a reduction semantics and barb predicates. It was
initially proposed by Milner and Sangiorgi for
CCS~\cite{Milner92barbed}, and adapted to many other process
calculi~\cite{Honda95reductionbased,Amadio96bisimulations}, including
the join calculus. We take \emph{weak
  barbed congruence}~\cite{Milner92barbed} as our basic notion of
behavioral equivalence for closed processes.

\subsubsection{Observational equivalence for closed processes}
\begin{defi}[Barb predicates]
  Let $P$ be a closed process, and $x$ be a free channel name in $P$,
  \begin{enumerate}[(1)]
  \item $P$ has a strong barb on channel $x$: $\barb{P}{x}$, iff $ P
    \equiv \define{D}{\para{Q}{x(\exp)}}$, for some $D$, $Q$ and
    $\exp$, where $x \not\in \dv{D}$.
  \item $P$ has a weak barb on channel $x$: $\weakbarb{P}{x}$, iff $P
    \reduces^* P'$, such that $\barb{P'}{x}$.
  \end{enumerate}
  where $\reduces^*$ denotes the reflexive and
  transitive closure of $\reduces$.
\end{defi}
Following the definition, it is easy to check that two structurally
congruent processes maintain the same barbs, \ie the lemma below.
\begin{lem}\label{lemma.barb}
  For two closed processes $P$ and $Q$, whenever $P\equiv Q$, we have
  $\barb{P}{x}$ iff $\barb{Q}{x}$, and $\weakbarb{P}{x}$ iff
  $\weakbarb{Q}{x}$.
\end{lem}
\begin{proof}
  The part for strong barb holds following the transitivity of
  $\equiv$, and the part for weak barb holds following
  Lemma~\ref{struct-rule}.  \forget{\qed}
\end{proof}

\begin{defi}[Weak barbed bisimulation]
\label{def.joinpi.bisi}
  A binary relation $\R$ on closed processes is a weak barbed
  bisimulation, iff whenever $P\R Q$, we have:
  \begin{enumerate}[(1)]
  \item If $P\reduces^* P'$, then $\exists Q'$, such that $Q\reduces^*
    Q'$ and $P'\R Q'$, and vice versa. ($\R$ is a reduction
    bisimulation.)
  \item $\weakbarb{P}{x}$ implies $\weakbarb{Q}{x}$ for any channel
    $x$, and vice versa. ($\R$ preserves barbs.)
  \end{enumerate}
\end{defi}
To make the definition easier to work with, we prove the following
lemma where $P\reduces^* P'$ is replaced by $P\reduces P'$ in the
first clause, and $\weakbarb{P}{x}$ is replaced by $\barb{P}{x}$ in
the second clause.
\begin{lem}\label{lemma.joinpi.bisi}
  Let $\R$ be a binary relation on closed processes that satisfies the
  following two conditions for any processes $P$ and $Q$ such that
  $P \R Q$:
  \begin{enumerate}[\em(1)]
  \item If $P\reduces P'$, then $\exists Q'$, such that $Q\reduces^*
    Q'$ and $P'\R Q'$, and vice versa.
  \item $\barb{P}{x}$ implies $\weakbarb{Q}{x}$ for any channel $x$,
    and vice versa.
  \end{enumerate}
  Then $\R$ is a weak barbed bisimulation.
\end{lem}
\begin{proof} We check against the two clauses of
  Definition~\ref{def.joinpi.bisi} for one direction. The proof of the
  other direction is symmetric.
  \begin{enumerate}[(1)]
  \item $\R$ is a reduction bisimulation, that is
    $$P\reduces^* P' \implies \exists Q', \mbox{ s.t. } Q\reduces^* Q'
    \mbox{ and } P'\R Q'$$ We reason on the length of the derivation
    $P\reduces^* P'$, written $n$.

    \paragraph{\bf Base case}\ $n = 0, 1$, trivial.

    \paragraph{\bf Induction case}\  As illustrated in the following
      diagram chase,
    \begin{center}
      \input{wbbisi-alt.fex}
    \end{center}
    we have $P \reduces^{n-1} P_1 \reduces P'$.  By induction
    hypothesis, we have $\exists Q_1, \mbox{ s.t. } Q\reduces^* Q_1
    \mbox{ and } P_1\R Q_1$. By applying hypothesis~$(1)$ to
    to $P_1$ and $Q_1$, we also have $\exists Q', \mbox{ s.t.  }
    Q_1\reduces^* Q' \mbox{ and } P'\R Q'$. And we conclude.
  \item $\R$ preserves barbs, that is 
    $\weakbarb{P}{x} \implies \weakbarb{Q}{x}$.
    We thus assume $\weakbarb{P}{x}$. That is,
    $$
    \exists P', P \reduces^* P' \mbox{ and } \barb{P'}{x}
    $$
    By $(1)$~above,
    $$
    \exists Q', Q\reduces^* Q' \mbox{ and } P'\R Q'
    $$
    Then by applying hypothesis~$(2)$ to $P'$ and~$Q'$,
    we get $\weakbarb{Q'}{x}$. Hence we have $\weakbarb{Q}{x}$.
  \end{enumerate}
\forget{\qed} 
\end{proof}
In later discussion, we sometimes directly check against the two
conditions of Lemma~\ref{lemma.joinpi.bisi} instead of the ones
of Definition~\ref{def.joinpi.bisi} for weak barbed bisimulation.

We define \emph{a context} as a term built by the grammar of process
with a single process placeholder $[\cdot]$.  \emph{An evaluation
  context} $\econtext{\cdot}$ is a context in which the placeholder is
not guarded. Namely: $$
\econtext{\cdot} \defineas [\cdot] \;\mid\;
\para{\econtext{\cdot}}{P} \;\mid\; \para{P}{\econtext{\cdot}}
\;\mid\; \define{D}{\econtext{\cdot}}\\
$$
In addition to evaluation contexts,
there are two kinds of \emph{guarded contexts},
referred to as \emph{definition contexts} (\ie
$\jctxt[P]{[\cdot]}$) and \emph{pattern matching contexts} (\ie
$\mctxt{[\cdot]}$).
We say that a context is closed if all the free variables in it are of
channel types.
\begin{defi}[Weak barbed congruence]
  A binary relation on closed processes is a weak barbed congruence, iff
  it is a weak barbed bisimulation and closed by application of any
  closed evaluation context. We denote the largest weak barbed
  congruence as~$\wbc$.
\end{defi}
The weak barbed congruence $\wbc$ is defined on the closed subset of
the applied join calculus. Although the definition itself only
requires the closure of evaluation contexts, it can be proved that the
full congruence does not provide more discriminative power. Similarly
to what Fournet has established for the pure join calculus in his
thesis~\cite{Fournet98:PhD}, we first have the property that $\wbc$ is
closed by substitution because, roughly, name substitutions may be
mimicked by evaluation contexts with ``forwarders''.
\begin{lem}\label{lemma.wbcsub}
  Given two closed processes $P$ and $Q$, if $P \wbc Q$, then for any
  substitution $\sigma$, $\sub{P}{\sigma}\wbc\sub{Q}{\sigma}$.
 (Note that ``closed'' stands for ``variable-closed''.)
\end{lem}
\begin{proof}
  The main idea is to build an evaluation context $\econtext{\cdot}$
  whose task is to forward messages from names to names
  according to the substitution $\sigma$, and to prove the
  equivalences $P\sigma \wbc \econtext{P}$ and $Q\sigma \wbc
  \econtext{Q}$. Because $\wbc$ is closed by evaluation contexts, we
  also have $P\wbc Q \implies \econtext{P}\wbc\econtext{Q}$. Then we
  conclude by the transitivity of $\wbc$. 
  Refer to the proof of Fournet in~\cite[Lemma 4.14 of Chapter
  4]{Fournet98:PhD} for details. \forget{\qed}
\end{proof}
Then based on this property, the full congruence is also
guaranteed considering the fact that the essence of a guarded
  context is substitution.
\begin{thm}\label{th.wbcfull}
  Weak barbed congruence $\wbc$ is closed by application of any
  closed context.
\end{thm}
\begin{proof}
  Corollary of
  Theorem~\ref{th.pseq-cong} that we prove later on. \forget{\qed}
\end{proof}

Up to now, we have defined the weak barbed congruence to express the
equivalence of two closed processes. However, our purpose is to study
the correctness of a static transformation. Since static
transformations apply perfectly well to processes with free
variables of non-channel type, restricting ourselves to the world of
closed processes is not an option. In the next section, we will derive
an equivalence relation for open processes. But before getting into
the definition, let us first establish some up-to techniques on the
closed sub-set of the calculus. Such up-to techniques will be used
during the courses of proving upcoming lemmas and theorems.

\begin{defi}[Weak barbed congruence up to $\equiv$]
\label{def.wbc-upto-equiv}
A binary relation $\R$ on closed processes is a weak barbed congruence
up to $\equiv$, iff $P\R Q$ implies:
  \begin{enumerate}[(1)]
  \item for any closed evaluation context $\econtext{\cdot}$,
    $\econtext{P} \equiv\R \equiv\econtext{Q}$ ($\R$ is closed
    under evaluation contexts up to $\equiv$);
  \item whenever $P \reduces^{*} P'$, $\exists Q'$, such that $Q
    \reduces^* Q'$ and $P' \equiv\R\equiv Q'$, and vice versa ($\R$ is
    a reduction bisimulation up to $\equiv$);
  \item $\weakbarb{P}{x}$ implies $\weakbarb{Q}{x}$ for any channel
    $x$, and vice versa. ($\R$ preserves barbs.)
  \end{enumerate}
\end{defi}
As we did for plain weak barbed bisimulation
(Definition~\ref{def.joinpi.bisi})
in Lemma~\ref{lemma.joinpi.bisi}, we introduce the following
weakened conditions for checking weak barbed congruence up to~$\equiv$.
\begin{lem}\label{lemma.wbc-upto-equiv-alt}
  Let $\R$ be a binary relation on closed processes and $\R$ that satisfies the
  following three conditions for any processes $P$ and $Q$ such that
  $P\R Q$:
  \begin{enumerate}[\em(1)]
  \item for any closed evaluation context $\econtext{\cdot}$,
    $\econtext{P} \equiv\R \equiv\econtext{Q}$;
  \item If $P\reduces P'$, then $\exists Q'$, such that $Q\reduces^*
    Q'$ and $P'\equiv\R\equiv Q'$, and vice versa.
  \item $\barb{P}{x}$ implies $\weakbarb{Q}{x}$ for any channel $x$,
    and vice versa.
  \end{enumerate}
  Then $\R$ is a weak barbed congruence up to $\equiv$.
\end{lem}
\begin{proof}
  We check against the three clauses of
  Definition~\ref{def.wbc-upto-equiv}.
  \begin{enumerate}[(1)]
  \item The first clause is the same as clause $(1)$ of
    Definition~\ref{def.wbc-upto-equiv}.
  \item We show: $$P\reduces^* P' \implies \exists Q', \mbox{ s.t. }
    Q\reduces^* Q' \mbox{ and } P'\equiv\R\equiv Q'$$ We reason on the
    length of the derivation $P\reduces^* P'$, written $n$.

    \paragraph{\bf Base case}\ $n = 0, 1$, trivial.

    \paragraph{\bf Induction case}\ As illustrated in the following
      diagram chase,
    \begin{center}
      \input{wbcuptoequiv-alt.fex}
    \end{center}
    we have $P \reduces^{n-1} P_1 \reduces P'$. By induction
    hypothesis, we have $\exists Q_1, Q_2, P_2$, s.t. $Q\reduces^* Q_1
    \mbox{ and } P_1\equiv P_2\R Q_2\equiv Q_1$.  Following
    Lemma~\ref{struct-rule}, we have $P_2\reduces P'$, too.  By
    applying hypothesis~$(2)$ to $P_2$ and $Q_2$, we also have
    $\exists Q', \mbox{ s.t.  } Q_2\reduces^* Q' \mbox{ and }
    P'\equiv\R\equiv Q'$. Then by Lemma~\ref{struct-rule} again, we
    have $Q_1\reduces^{*} Q'$, too. To conclude, we have $\exists Q',
    \mbox{ s.t. } Q\reduces^* Q' \mbox{ and } P'\equiv\R\equiv Q'$.

    The proof of the other direction is symmetric.
  \item We show: $$\weakbarb{P}{x} \implies \weakbarb{Q}{x}$$
    We thus assume $\weakbarb{P}{x}$:
    $$
    \exists P_1, \mbox{ s.t. }P\reduces^{*} P_1 \mbox{ and } \barb{P_1}{x}
    $$
    By $(2)$~above, we get:
    $$
    \exists Q_1, Q_2, P_2, \mbox{ s.t. } Q\reduces^* Q_1\mbox{ and }
    P_1 \equiv P_2 \R Q_2 \equiv Q_1
    $$
    By Lemma~\ref{lemma.barb}, we have $\barb{P_2}{x}$. Applying
    hypothesis~$(3)$ to $P_2$ and~$Q_2$, we get
    $\weakbarb{Q_2}{x}$. Then by Lemma~\ref{lemma.barb} again, we have
    $\weakbarb{Q_1}{x}$. To conclude, we have $Q\reduces^* Q_1$ and
    $\weakbarb{Q_1}{x}$, \ie $\weakbarb{Q}{x}$.
    The proof of the other direction is symmetric.
  \end{enumerate}
\end{proof}

\begin{lem}\label{lemma.wbc-upto-equiv}
  If $\R$ is a weak barbed congruence up to $\equiv$, then
  $\R\,\subseteq\,\wbc$.
\end{lem}
\begin{proof}
  We first show $\equiv\R\equiv$ $\subseteq$ $\wbc$, \ie $\equiv\R\equiv$ is
  a weak barbed congruence.
  \begin{enumerate}[(1)]
  \item $\equiv\R\equiv$ is closed under evaluation contexts. Given
    $P\equiv\R\equiv Q$, there exist $P_1$ and $Q_1$ such that
    $P\equiv P_1\R Q_1\equiv Q$.
    Let us name two properties:
    \begin{enumerate}
    \item $\equiv$ is closed under evaluation contexts;
    \item clause $(1)$ of Definition~\ref{def.wbc-upto-equiv}.
    \end{enumerate}
    Then, for any closed evaluation context $\econtext{\cdot}$, we have:
    $$
    \econtext{P} \stackrel{(a)}\equiv \econtext{P_1} 
    \overbrace{\equiv\R\equiv}^{(b)} 
    \econtext{Q_1} \stackrel{(a)}\equiv \econtext{Q}
    $$
    By transitivity of~$\equiv$, we conclude:
    $$
    \econtext{P} \equiv\R\equiv \econtext{Q}
    $$
  \item $\equiv\R\equiv$ is a reduction bisimulation. We use
     clause~$(2)$ of Definition~\ref{def.wbc-upto-equiv} and then
     Lemma~\ref{struct-rule} to reason by
     diagram chase as follows:
    \begin{center}
      \input{wbcuptoequiv.fex}
    \end{center}
  \item $\equiv\R\equiv$ preserves barbs. Given $P\equiv\R\equiv Q$,
    we have $P\equiv P_1\R Q_1\equiv Q$, and the following statement,
    $$
    \barb{P}{x}
    \stackrel{Lemma~\ref{lemma.barb}}\Longrightarrow
    \barb{P_1}{x}
    \stackrel{Def~\ref{def.wbc-upto-equiv}.(3)}\Longrightarrow
    \weakbarb{Q_1}{x}
    \stackrel{Lemma~\ref{lemma.barb}}\Longrightarrow
    \weakbarb{Q}{x}
    \mbox{, and vice versa.}
    $$
  \end{enumerate}
  Then because $\R$ $\subseteq$ $\equiv\R\equiv$ $\subseteq$ $\wbc$,
  we conclude that $\R$ $\subseteq$ $\wbc$. \forget{\qed}
\end{proof}

A standard proof technique is then to consider weak barbed congruence up
to~$\wbc$.
However, as demonstrated in~\cite{SaMi92}, such a technique does not work
in general in weak settings.
Thus, we instead define another relation, where up to $\wbc$
is performed on one side only.
This new relation is sound, as shown by the forthcoming
Lemma~\ref{lemma.wbc-upto-id}.

\begin{defi}[Weak barbed congruence up to \Id~\footnote{\Id stands for
    the identity relation on closed processes. Note that this relation
    is derived from  ``bisimulation up to almost-weak
    bisimulation'' in~\cite{SaMi92}, because \Id is included in
    almost-weak bisimulation, with some adjustments to the
    barbed setting.}]
\label{def.wbc-upto-id}
  A binary relation~$\R$ on closed processes is a weak barbed
  congruence up to \Id, iff $P\R Q$ implies:
  \begin{enumerate}[(1)]
  \item for any closed evaluation context $\econtext{\cdot}$,
    $\econtext{P} \equiv\R \equiv\econtext{Q}$ ($\R$ is closed under
    evaluation contexts up to $\equiv$);
  \item whenever $P \reduces^{*} P'$, $\exists Q'$, such that $Q
    \reduces^* Q'$ and $P' \R\wbc Q'$;
  \item whenever $Q \reduces^{*} Q'$, $\exists P'$, such that $P
    \reduces^* P'$ and $P' \wbc\R Q'$;
    
    (The two clause above say that $R$ is a reduction bisimulation up to
    \Id.)
  \item $\weakbarb{P}{x}$ implies $\weakbarb{Q}{x}$ for any channel
    $x$, and vice versa. ($\R$ preserves barbs.)
  \end{enumerate}
\end{defi}
Again, we first derive the following alternative conditions for
checking weak barbed congruence up to \Id.
\begin{lem}\label{lemma.wbc-upto-id-alt}
  $\R$ is a binary relation on closed processes and $\R$ satisfies the
  following conditions for any processes $P$ and $Q$ such that $P\R
  Q$:
  \begin{enumerate}[(1)]
  \item for any closed evaluation context $\econtext{\cdot}$,
    $\econtext{P} \equiv\R \equiv\econtext{Q}$;
  \item whenever $P \reduces P'$, $\exists Q'$, such that $Q
    \reduces^* Q'$ and $P' \R\wbc Q'$;
  \item whenever $Q \reduces Q'$, $\exists P'$, such that $P
    \reduces^* P'$ and $P' \wbc\R Q'$;
  \item $\barb{P}{x}$ implies $\weakbarb{Q}{x}$ for any channel
    $x$, and vice versa.
  \end{enumerate}
  Then $\R$ is a weak barbed congruence up to \Id.
\end{lem}
\begin{proof}
  We check against the clauses of Definition~\ref{def.wbc-upto-id}.
  \begin{enumerate}
  \item The first clause is the same.
  \item We show: $$P\reduces^* P' \implies \exists Q', \mbox{ s.t. }
    Q\reduces^* Q' \mbox{ and } P'\R\wbc Q'$$ We reason on the
    length of the derivation $P\reduces^* P'$, written $n$.

    \paragraph{\bf Base case}\ $n=0, 1$, trivial.

    \paragraph{\bf Induction case}\ As illustrated in the following
      diagram chase, 
      \begin{center}
        \input{redbisiuptoid.fex}
      \end{center}
      we have $P \reduces^{n-1} P_1 \reduces P'$. By induction
      hypothesis, we get $\exists Q_1$, such that $Q \reduces^* Q_1$
      and $P_1 \R\wbc Q_1$. That is, $\exists Q_2$, such that $P_1 \R
      Q_2 \wbc Q_1$.  By applying hypothesis~$(2)$ to $P_1$ and~$Q_2$,
      we have $\exists Q_3$ such that $Q_2 \reduces^* Q_3$ and
      $P'\R\wbc Q_3$.  Because $Q_2 \wbc Q_1$, we also have $\exists
      Q'$ such that $Q_1 \reduces^* Q'$ and $Q_3 \wbc Q'$ --- remember
      that $\wbc$ is the largest weak barbed congruence and thus a
      reduction bisimulation.  We conclude by transitivity of~$\wbc$.
  \item Symmetric of (2) above.
  \item We show: $$\weakbarb{P}{x} \implies \weakbarb{Q}{x}$$
    We thus assume $\weakbarb{P}{x}$. That is, we have:
    $$
    \exists P_1\mbox{ s.t }P\reduces^{*} P_1 \mbox{ and } \barb{P_1}{x}
    $$
    By $(2)$~above, we get:
    $$
    \exists Q_1, Q_2, \mbox{ s.t. } Q\reduces^* Q_1 \mbox{ and }
    P_1\R Q_2\wbc Q_1
    $$
    Applying hypothesis~$(4)$ to $P_1$ and~$Q_2$, we get
    $\weakbarb{Q_2}{x}$. Applying clause $(2)$ of
    Definition~\ref{def.joinpi.bisi} to $Q_2$ and $Q_1$, we then get
    $\weakbarb{Q_1}{x}$. To conclude, we have $Q\reduces^* Q_1$ and
    $\weakbarb{Q_1}{x}$, \ie $\weakbarb{Q}{x}$.  The proof of the
    other direction is symmetric.
  \end{enumerate}
\end{proof}

\begin{lem}\label{lemma.wbc-upto-id}
  If $\R$ is a weak barbed congruence up to \Id, then
  $\R\,\subseteq\,\wbc$.
\end{lem}
\begin{proof}
  We first show $\wbc\R\wbc$ $\subseteq$ $\wbc$, \ie $\wbc\R\wbc$ is
  a weak barbed congruence.
  \begin{enumerate}[(1)]
  \item $\wbc\R\wbc$ is closed under evaluation contexts. Given
    $P\wbc\R\wbc Q$, there exist $P_1$ and $Q_1$ such that $P\wbc
    P_1\R Q_1\wbc Q$. Let us name two properties:
    \begin{enumerate}[(a)]
    \item $\wbc$ is closed under evaluation contexts;
    \item clause $(1)$ of Definition~\ref{def.wbc-upto-id};
    \end{enumerate}
    Then, for any closed evaluation context $\econtext{\cdot}$, we have:
    $$
    \econtext{P} \stackrel{(a)}\wbc \econtext{P_1} 
    \overbrace{\equiv\R\equiv}^{(b)} 
    \econtext{Q_1} \stackrel{(a)}\wbc \econtext{Q}
    $$
    Because $\equiv$ $\subseteq$ $\wbc$, we have $\equiv\R\equiv$
    $\subseteq$ $\wbc\R\wbc$. Hence we have:
    $$
    \econtext{P} \wbc \econtext{P_1} 
    \wbc\R\wbc 
    \econtext{Q_1} \wbc \econtext{Q}
    $$
    And we conclude, by transitivity of~$\wbc$.
  \item $\wbc\R\wbc$ is a reduction bisimulation. We use clause~$(1)$
    of Definition~\ref{def.joinpi.bisi}, clause $(2)$ of
    Definition~\ref{def.wbc-upto-id}, clause $(1)$ of
    Definition~\ref{def.joinpi.bisi}, and the transitivity of $\wbc$,
    in the proof sketched by the following diagram:
    \begin{center}
      \input{wbcuptoid.fex}
    \end{center}
  \item $\wbc\R\wbc$ preserves barbs. Given $P\wbc\R\wbc Q$, we have
    $P\wbc P_1\R Q_1\wbc Q$, and the following statement,
    $$
    \barb{P}{x}
    \stackrel{Def~\ref{def.joinpi.bisi}.(2)}\Longrightarrow
    \weakbarb{P_1}{x}
    \stackrel{Def~\ref{def.wbc-upto-id}.(3)}\Longrightarrow
    \weakbarb{Q_1}{x} \stackrel{Def~\ref{def.joinpi.bisi}.(2)}
    \Longrightarrow \weakbarb{Q}{x}
    \mbox{, and vice versa.}
    $$
  \end{enumerate}
  Then because $\R$ $\subseteq$ $\wbc\R\wbc$ $\subseteq$ $\wbc$,
  we conclude that $\R$ $\subseteq$ $\wbc$.  \forget{\qed}
\end{proof}

\subsubsection{Observational equivalence for open processes}
The approach we follow here is to lift the equivalence relation of
closed processes to open processes by closing up by all substitutions,
and we call the resulting relation \emph{open equivalence}.

Although both are ``open'', our open equivalence is
unrelated to the open bisimilarity of Sangiorgi
in~\cite{Sangiorgi96OpenBisi}. We use ``open'' to name our equivalence
relation because it relates
open terms. By contrast, ``open'' in open
bisimilarity emphasizes a characteristic of the bisimulation
definition, namely free names are open to equality throughout the
bisimulation game. From the perspective of where and when to apply
name substitutions, for open equivalence, we instantiate free names
(and variables) only at the beginning before we test the resulting (closed)
processes for weak barbed congruence.
On the contrary, in the case of open bisimilarity, such instantiation
happens at every co-inductive step.

Another way to define equivalence relations on open terms could be to
adapt the semantics to symbolic transition system and to define a
symbolic barbed congruence like in~\cite{SymbolicBisi}.  Although the
symbolic method is claimed to be easier for analysis and verification,
we found open equivalence to be lighter and more intuitive.  As a
matter of fact, it is not uncommon to define functions extensionally,
\ie by considering application to all possible arguments.  Moreover,
as can be seen in Section~\ref{sec.correct}, our proofs remain
tractable.
\begin{defi}[Open equivalence $\pseq$]
  \label{def.joinpi.pseq}
  Two processes $P$ and $Q$ are open equivalent, written $P\pseq Q$,
  iff for any substitution $\sigma$ such that ${P}{\sigma}$ and
  ${Q}{\sigma}$ are closed, we have ${P}{\sigma} \wbc {Q}{\sigma}$.
\end{defi}
As a corollary, $\pseq$ is closed by any substitution.
\begin{lem}\label{lemma.pseq-sub}
  $P \pseq Q \implies \forall \sigma. \sub{P}{\sigma} \pseq
  \sub{Q}{\sigma}$
\end{lem}
\resetequationcounter
\begin{proof}
  We assume $P \pseq Q$ and let $\sigma$ be a substitution. We need to
  prove that $P\sigma \pseq Q\sigma$. That is,
  we need to prove that, for all closing substitution~$\rho$, we
  have:
\begin{equation*}
{({P}{\sigma})}{\rho} \wbc
    {({Q}{\sigma})}{\rho}
\end{equation*}
Thus, we need to prove that, for all closing substitution~$\rho$, we
have:
\begin{equation}
 {P}{(\composite{\rho}{\sigma})} \wbc
  {Q}{(\composite{\rho}{\sigma})} \label{lb.joinpi.pseq-subb}
\end{equation}
where $\composite{}{}$ stands for substitution composition, \ie
$P(\composite{\rho}{\sigma})\defineas (P\sigma)\rho$.
It remains to observe that
$\composite{\rho}{\sigma}$ closes both processes $P$ and~$Q$,
and to apply the definition of~$P \pseq Q$, before concluding
that statement~(\ref{lb.joinpi.pseq-subb}) above holds. \forget{\qed}
\end{proof}

We aim at proving that $\pseq$ is closed by any contexts
(Theorem~\ref{th.pseq-cong} below). To prove the theorem, we need the
following rather unusual lemma, to state the fact that although we
have introduced ``deterministic'' reduction into the process calculus
by extending it with the \kwd{match} construct, this kind of
determinism does not impact process equivalence.
\begin{lem}\label{lemma.joinpi.reduces}
  We say a closed process $P$ \emph{deterministically reduces} to $P'$,
  iff for all $P''$ such that $P\reduces P''$, we have $P'\equiv P''$.
  For any such pair of closed processes $P$ and $P'$, we have $P \wbc
  P'$ .
\end{lem}
\begin{proof}
  Let $\R$ be the relation $\set{(\define{D}{\para{P}{Q}},
  \;\define{D}{\para{P'}{Q}}), (S,S)}$ for all closed definitions
  $D$, closed processes $Q$ and $S$, and all $(P, P')$ pairs such
  that $P$ deterministically reduces to~$P'$. We prove that $\R$
  is a weak barbed congruence up to~$\equiv$.
  \begin{enumerate}[$\bullet$]
  \item By definition, $\R$ is closed by evaluation contexts
    up to $\equiv$ (\ie Lemma~\ref{lemma.wbc-upto-equiv-alt}.(1)).
  \item We show that $\R$ preserves barbs (\ie
    Lemma~\ref{lemma.wbc-upto-equiv-alt}.(3)). We omit the (trivial) discussion
    of pairs of identical processes $(S,S)$ in $\R$.
    We show that $\barb{(\define{D}{\para{P}{Q}})}{x} \implies
    \weakbarb{(\define{D}{\para{P'}{Q}})}{x}$. We distinguish the
    cases that make $\barb{(\define{D}{\para{P}{Q}})}{x}$ hold.
    \begin{enumerate}[$-$]
    \item $\barb{P}{x}$.
      Obviously reduction cannot erase a barb  ($x\not\in\dv{D}$), \ie
      we have $\barb{P'}{x}$.
      Hence, we have $\barb{(\define{D}{\para{P'}{Q}})}{x}$.
    \item $\barb{Q}{x}$. Trivial.
    \end{enumerate}
    As to the opposite direction \ie
    $\barb{(\define{D}{\para{P'}{Q}})}{x} \implies
    \weakbarb{(\define{D}{\para{P}{Q}})}{x}$, it holds trivially
    because $\define{D}{\para{P}{Q}} \reduces
    \define{D}{\para{P'}{Q}}$.
  \item We show $\R$ to be a reduction bisimulation up to $\equiv$
    (\ie Lemma~\ref{lemma.wbc-upto-equiv-alt}.(2)). We omit the
    trivial case of pairs of identical processes in $\R$, that is,
    we only consider process pairs of form: $(\define{D}{\para{P}{Q}},
    \;\define{D}{\para{P'}{Q}})$.
    \begin{enumerate}[$-$]
    \item If the reduction of the left part is caused by a reduction
      on $Q$ alone or by the interaction between $D$ and $Q$, yielding
      $\define{D}{\para{P}{Q'}}$, then the right part can perform
      the same reduction step, yielding $\define{D}{\para{P'}{Q'}}$.
      The resulting two processes are still in relation $\R$ with $Q$
      being $Q'$. Vice versa.
    \item If the reduction of the left part is caused by a reduction
      on $P$ alone, then, because $P$ deterministically reduces to $P'$,
      the resulting process
      is $\define{D}{\para{P'}{Q}}$ (up to $\equiv$).
      Thus, the right part simulates
      with no reduction and $\define{D}{\para{P'}{Q}}$ satisfies
      relation $\R$ with itself.
    \item If the reduction of the left part is caused by the
      interaction between $D$ and $P$, then we must have $P \equiv
      \para{P_0}{{J}{\sigma}}$ where $\reaction{J}{G}$ is a reaction
      rule in $D$ and the resulting process is
      $\define{D}{\para{P_0}{\para{Q}{G\sigma}}}$. Because $J\sigma$
      does not reduce by itself and $P$ deterministically reduces to
      $P'$, we have $P' \equiv \para{P'_0}{{J}{\sigma}}$ and $P_0$
      deterministically reduces to $P'_0$. Therefore, the right part
      simulates by an identical reduction and gives
      $\define{D}{\para{P'_0}{\para{Q}{G\sigma}}}$. The resulting two
      processes are still in relation $\R$ with $Q$ being
      $\para{Q}{G\sigma}$, $P$ being $P_0$, and $P'$ being $P'_0$.
    \item If the reduction of the right part is caused by a reduction
      on $P'$ itself or by the interaction between $D$ and $P'$, then
      the left part can always simulate the reduction by first
      reducing $\define{D}{\para{P}{Q}}$ to
      $\define{D}{\para{P'}{Q}}$.
    \end{enumerate}
  \end{enumerate}
  Following the analysis above, $\R$ is a weak barbed congruence up to
  $\equiv$. Besides we have $$P \equiv
  (\define{\top}{\para{P}{\nullp}})\, \R \,
  (\define{\top}{\para{P'}{\nullp}})\equiv P'$$
  Moreover, by the proof of Lemma~\ref{lemma.wbc-upto-equiv},
  relation $\equiv \R \equiv$ is a weak barbed congruence. Hence we conclude
  $P \wbc P'$. \forget{\qed}
\end{proof}

\begin{thm}\label{th.pseq-cong}
  The open equivalence $\pseq$ is a full congruence.
\end{thm}
\resetequationcounter
\begin{proof}
  We demonstrate $\pseq$ is closed by 1. evaluation contexts, 2.
  definition contexts, and 3. pattern matching contexts.
  In the proof, we locally use $A$, $B$, $R$, $S$, $T$, $V$, $W$, $X$,
  $Y$, $Z$ to denote various processes.

  \paragraph{\bf 1. Closed by evaluation contexts:
    $\econtext{\cdot}$.}  We show: $$P \pseq Q \implies
  \econtext{P}\pseq \econtext{Q}$$ For any substitution $\sigma$ such
  that ${(\econtext{P})}{\sigma}$ and ${(\econtext{Q})}{\sigma}$ are
  closed, we need to prove ${(\econtext{P})}{\sigma} \wbc
  {(\econtext{Q})}{\sigma}$.  We write ${(\econtext{P})}{\sigma}$ as
  ${E}{\sigma}[{P}{\sigma_1}]$ and ${(\econtext{Q})}{\sigma}$ as
  ${E}{\sigma}[{Q}{\sigma_1}]$, where ${E}{\sigma}[\cdot]$,
  ${P}{\sigma_1}$, ${Q}{\sigma_1}$ are closed and $\sigma_1$ is
  $\sigma$ minus the (possible) bindings for the channel names bound
  by $E$ in $[\cdot]$. By hypothesis $P \pseq Q$, we have
  ${P}{\sigma_1} \wbc {Q}{\sigma_1}$.  Then, ${E}{\sigma}[\cdot]$
  being a closed evaluation context, we conclude, by definition
  of~$\wbc$.
  
  \paragraph{\bf 2. Closed by definition contexts: $\jctxt{[\cdot]}$.}
  We show: $$P \pseq Q \implies (\jctxt{P}) \pseq (\jctxt{Q})$$ For
  any substitution $\sigma$ such that ${(\jctxt{P})}{\sigma}$ and
  ${(\jctxt{Q})}{\sigma}$ are closed, we need to prove: $$
  {(\jctxt{P})}{\sigma} \wbc {(\jctxt{Q})}{\sigma} $$ namely,
  \begin{align}
    \define{\dis{\reaction{J}{{P}{\sigma_1}}}{{D}{\sigma_2}}}
    {{R}{\sigma_2}}
    \wbc
    \define{\dis{\reaction{J}{{Q}{\sigma_1}}}{{D}{\sigma_2}}}
    {{R}{\sigma_2}} \label{lb.joinpi.pseq-cong1}
  \end{align}
  where $\sigma_2$ is $\sigma$ minus the (possible) bindings for the
  channel names defined in $\dis{\reaction{J}{P}}{D}$ (\ie
  $\dv{\dis{\reaction{J}{P}}{D}}$), and $\sigma_1$ is $\sigma_2$ minus
  the (possible) bindings for the variables of~$\rv{J}$.  Notice that,
  by contrast with the subcomponents $D\sigma_2$ and~$R\sigma_2$ that
  are closed, the processes $P\sigma_1$ and $Q\sigma_1$ may \emph{not}
  be closed, since some of the variables in~$\rv{J}$ may be of an
  algebraic type.  Nevertheless, by hypothesis $P \pseq Q$ and
  Lemma~\ref{lemma.pseq-sub}, we have ${P}{\sigma_1} \pseq
  {Q}{\sigma_1}$.

  Then, we build the following relation $\R$ on closed processes:
  $$
  \R = \set{(\define{\dis{\reaction{J}{S}}{D}}{A},
    \define{\dis{\reaction{J}{T}}{D}}{B}) \mid S\pseq T \mbox{ and } A \wbc
    B}.$$
  We analyze the following three aspects of $\R$: closure
  by closed evaluation contexts; preserving barbs; and reduction
  bisimulation.

  \begin{enumerate}[$\bullet$]
  \item $\R$ is closed by closed evaluation contexts up to $\equiv$
    (\ie Lemma~\ref{lemma.wbc-upto-id-alt}.(1)). For any closed
    $\econtext{\cdot}$, with necessary $\alpha$-conversions left
    implicit, we have:
    \begin{align*}
      \econtext{\define{\dis{\reaction{J}{S}}{D}}{A}} & \equiv
      \define{\dis{\reaction{J}{S}}{(\dis{D}{D'})}}{(\para{A}{K})}
      \\
      \econtext{\define{\dis{\reaction{J}{T}}{D}}{B}} & \equiv 
      \define{\dis{\reaction{J}{T}}{(\dis{D}{D'})}}{(\para{B}{K})}
    \end{align*}
    where $\para{A}{K} \wbc \para{B}{K}$, because $\wbc$ is preserved by
    the closed evaluation context $\para{[\cdot]}{K}$.
   
    \def\C#1{\dd[#1]}
  \item $\R$ preserves barbs (\ie
    Lemma~\ref{lemma.wbc-upto-id-alt}.(4)).  We write $\C{X, Y}$ for
    the closed process $\define{\dis{\reaction{J}{X}}{D}}{Y}$.  Since
    $\R$ is a symmetric relation, we only need to prove:
    $$\barb{\C{S,A}}{x} \implies \weakbarb{\C{T,B}}{x}$$
    Because $\barb{\C{S,A}}{x}$ implies $\barb{A}{x}$ and $x
    \not\in(\dv{J}\cup\dv{D})$, we also have $\barb{\C{T,A}}{x}$.
    Moreover, because $\C{T,\cdot}$ is a closed evaluation context,
    and by hypothesis $A \wbc B$, we have
    \begin{align}
      \C{T,A} &\wbc \C{T,B} \label{st.barb1}
    \end{align}
    By clause $(2)$ of Definition~\ref{def.joinpi.bisi}, we finally get
    $\weakbarb{\C{T,B}}{x}$.

  \item $\R$ is a reduction bisimulation up to \Id (\ie
    Lemma~\ref{lemma.wbc-upto-id-alt}.(2) and (3)).  We first prove
    the following statement.  For any two $\C{S,A}$ and~$\C{T,A}$, we
    have:
    \begin{align}
      \mbox{If $\C{S,A}\reduces W$, then $\C{T,A}\reduces V$, and
        $W\R V$.} \label{st.red2}
    \end{align}    
    There are three subcases, depending on the nature of the reduction
    to~$W$. 
\begin{enumerate}[(1)]
    \item $A\reduces A'$ and $W = \C{S,A'}$. Then
    $\C{T,A} \reduces \C{T,A'}$, with obviously $\C{S,A'} \R \C{T,A'}$,
    since $A'\wbc A'$.

    \item $A\equiv\para{A_0}{{J}{\eta}}$ and
    $W=\C{S,\para{A_0}{{S}{\eta}}}$. Then $\C{T,A} \reduces
    \C{T,\para{A_0}{{T}{\eta}}}$.
    Notice that ${S}{\eta}$ and ${T}{\eta}$ are
    closed. Then, from $S\pseq T$, we get ${S}{\eta} \wbc {T}{\eta}$,
    and thus  $\para{A_0}{{S}{\eta}} \wbc
    \para{A_0}{{T}{\eta}}$.  That is, we get $\C{S,\para{A_0}{{S}{\eta}}}
    \R \C{T,\para{A_0}{{T}{\eta}}}$.

    \item $A \equiv \para{A_0}{{J_i}{\eta_i}}$, $D$ has
    form $\ldots \kwd{or}\;\reaction{J_i}{P_i}\;\kwd{or}\ldots$, and $W
    =\C{S,\para{A_0}{{P_i}{\eta_i}}}$. Then $\C{T,A} \reduces
    \C{T,\para{A_0}{{P_i}{\eta_i}}}$. And we conclude,
     as we did in case 1 above.
\end{enumerate}\medskip

Moreover, from equivalence \eqref{st.barb1}
and since $\wbc$ is a bisimulation, we have:
    \begin{align}
      \label{st.red1}
      \mbox{If $\C{T,A}\reduces V$, then $\exists V'$ s.t. 
        $\C{T,B}\reduces^* V'$, and $V\wbc V'$.}
    \end{align}
    Combining both statements \eqref{st.red2} and~\eqref{st.red1},
    we get:
    \begin{align}
      \label{st.red}
      \mbox{If $\C{S,A}\reduces W$, then $\exists V'$ s.t.
        $\C{T,B}\reduces^* V'$, and $W \mathrel{\R\,\wbc} V'$.}
    \end{align}
    The proof of the other direction is by symmetry.
  \end{enumerate}

  Following the analysis above, $\R$ is a weak barbed congruence up to
  \Id, hence by Lemma~\ref{lemma.wbc-upto-id}, $\R\,\subseteq\,\wbc$.
  Obviously, the two processes of statement
  \eqref{lb.joinpi.pseq-cong1} are related by~$\R$. Therefore,
  \eqref{lb.joinpi.pseq-cong1} holds.  In other words, $\pseq$ is
  closed by any definition context.

  \paragraph{\bf 3. Closed by pattern matching contexts: $\mctxt{[\cdot]}$}
  We show: \begin{multline} P \pseq Q \implies \\ (\mctxt{P}) \pseq (\mctxt{Q})
  \end{multline}
  To establish the right part, we need to show: $$
  {(\mctxt{P})}{\sigma} \wbc {(\mctxt{Q})}{\sigma} $$
  for all $\sigma$, s.t. ${(\mctxt{P})}{\sigma}$ and ${(\mctxt{Q})}{\sigma}$
  are closed. Namely,
  \begin{align}
    \prefix{match} {\exp}{\sigma} \infix{with} \ldots \mid \pt_k
    \rightarrow {P}{\sigma_k} \mid \ldots \wbc \prefix{match}
    {\exp}{\sigma} \infix{with} \ldots \mid \pt_k \rightarrow
    {Q}{\sigma_k} \mid \ldots
    \label{lb.joinpi.pseq-cong2}
  \end{align}
  where $\sigma_k$ is $\sigma$ minus the (possible) bindings for the
  variables of $\rv{\pt_k}$. Notice that $e\sigma$ is closed, while
  $P\sigma_k$ and $Q\sigma_k$ may not be.

  By the semantics of ML pattern matching, $\prefix{match}
  {\exp}{\sigma} \infix{with} \ldots \mid \pt_k \rightarrow
  {P}{\sigma_k} \mid \ldots$ deterministically reduces to either
  ${P}{(\composite{\eta_k}{\sigma_k})}$ or ${R_i}{\eta_i}$, depending
  on the value of ${\exp}{\sigma}$.  Process $R_i$ is the $i$th
  guarded process ($i\neq k$) in this pattern matching, $\eta_k$ and
  $\eta_i$ stand for the substitutions that originate from algebraic
  matching.  Notice that ${P}{(\composite{\eta_k}{\sigma_k})}$ and
  ${R_i}{\eta_i}$ now are closed processes.  We have the similar
  statement for $\prefix{match}{\exp}{\sigma} \infix{with} \ldots \mid
  \pt_k \rightarrow {Q}{\sigma_k} \mid \ldots$.  Therefore, by
  Lemma~\ref{lemma.joinpi.reduces}, we have either:
  \begin{align}
    \label{st.matchl1}
    \prefix{match} {\exp}{\sigma}\infix{with}\ldots \mid\pt_k \rightarrow
    {P}{\sigma_k} \mid\ldots & \wbc
    {P}{(\composite{\eta_k}{\sigma_k})}    \\
    \label{st.matchr1}
    \prefix{match} {\exp}{\sigma} \infix{with}\ldots \mid\pt_k
    \rightarrow
    {Q}{\sigma_k} \mid\ldots & \wbc {Q}{(\composite{\eta_k}{\sigma_k})} \\
    \intertext{or we have:}
    \label{st.matchl2}
    \prefix{match} {\exp}{\sigma} \infix{with}\ldots \mid\pt_k
    \rightarrow {P}{\sigma_k} \mid\ldots & \wbc {R_i}{\eta_i}    \\
    \label{st.matchr2}
    \prefix{match}{\exp}{\sigma} \infix{with} \ldots \mid\pt_k \rightarrow
    {Q}{\sigma_k} \mid\ldots & \wbc {R_i}{\eta_i}
  \end{align}
  Obviously we have $R_i\eta_i \wbc R_i\eta_i$. Moreover, since $P
  \pseq Q$, we get ${P}{(\composite{\eta_k}{\sigma_k})} \wbc
  {Q}{(\composite{\eta_k}{\sigma_k})}$. Then, by the transitivity
  of~$\wbc$ and, either by \eqref{st.matchl1}--\eqref{st.matchr1}, or
  by \eqref{st.matchl2}--\eqref{st.matchr2}, we conclude that the
  statement~\eqref{lb.joinpi.pseq-cong2} holds.

  Additionally, in the case where $e\sigma$ matches none of the patterns
  in~\eqref{lb.joinpi.pseq-cong2}, both processes are blocked
  and are $\wbc$ to the null process $\nullp$. \forget{\qed}
\end{proof}

There is still a good property worth noticing: for the closed
subset of the applied join-calculus,  the equivalences $\pseq$ and $\wbc$
coincide. This is straightforward by the definition of~$\pseq$
and by Lemma~\ref{lemma.wbcsub}.
Then, Theorem~\ref{th.wbcfull} follows as a corollary.

\section{Transforming pattern arguments into ML pattern matching}
\label{sec.trans-idea}

The extension of the join calculus that we have presented up to now
remains quite  simple, in particular as regards chemical
semantics. However, an efficient implementation is more involved. Our
approach is to first transform the extended join definitions into
ordinary ones plus ML pattern matching, then reuse the existing
implementation of join. In this section, we explain informally the
key ideas of the transformation.

The extended join-pattern matching in applied join requires to test
message contents against pattern arguments, while the ordinary
join-pattern matching in join is only capable of testing message
presence.  Our idea is to separate algebraic pattern testing from
join-pattern synchronization, and to perform the former operation by
using ML pattern matching.  To avoid inappropriate message
consumption, message contents are tested first.  Let us consider the
following join definition where channel~\lst"x" has two pattern
arguments:
\begin{lstlisting}{Join}
def x($\pt_1$) & y$_1$($\ldots$) |> $P_1$
 or x($\pt_2$) & y$_2$($\ldots$) |> $P_2$
\end{lstlisting}
We refine channel~\lst"x" into more precise ones, each
of which carries the instances of patterns~$\pt_1$ or $\pt_2$:
\begin{lstlisting}{Join}
def x$_{\pt_1}$($\ldots$) & y$_1$($\ldots$) |> $P_1$
 or x$_{\pt_2}$($\ldots$) & y$_2$($\ldots$) |> $P_2$
\end{lstlisting}
Then, we add a new reaction rule to dispatch the messages on
channel~\lst"x" to either \lst"x"$_{\pt_1}$ or~\lst"x"$_{\pt_2}$:
\begin{lstlisting}{Join}
 or x(z) |> match z with
            | $\pt_1$ -> x$_{\pt_1}$(...)
            | $\pt_2$ -> x$_{\pt_2}$(...)
            | _ -> $\nullp$
\end{lstlisting}
Note that the null process is used in the last matching rule to discard
messages that match neither $\pt_1$ nor~$\pt_2$.

The simple compilation above works perfectly, as long as $\pt_1$ and
$\pt_2$ are incompatible. Unfortunately, it falls short when $\pt_1$
and $\pt_2$ have common instances.
Consider the situation where there is a message pending
on channel~$\id{y}_2$, none on~$\id{y}_1$, and also a message~$v$
on~$\id{x}$ where $v$ is a common instance of patterns $\pt_1$ and $\pt_2$.
Then, following the first match policy, the
deterministic ML pattern matching can only dispatch $\id{x}(v)$ to the
refined channel $\id{x}_{\pt_1}$.
As a result, the guarded process~$P_2$ is not
triggered, whereas it could have been.\footnote{Given
our implementation ``limited fairness guarantee'', it
can be argued that~$P_2$ should be~triggered.}
To tackle this problem, further
refinements are called for according to the following cases.
\begin{enumerate}[$\bullet$]
\item If $\pt_1 \preceq \pt_2$, (but $\pt_2 \not\preceq \pt_1$), that
  is if all instances of $\pt_2$ are instances of $\pt_1$, then, to
  get a chance of meeting its instances, pattern~$\pt_2$ must come
  first:
\begin{lstlisting}{Join}
 or x(z) |> match z with
           | $\pt_2$ -> x$_{\pt_2}$(...)
           | $\pt_1$ -> x$_{\pt_1}$(...)
           | _ -> $\nullp$
\end{lstlisting}
But now, channel $\id{x}_{\pt_1}$ does not carry all the possible
instances of pattern~$\pt_1$ any more, instances shared by
pattern~$\pt_2$ are dispatched to~$\id{x}_{\pt_2}$. As a consequence,
the actual transformation of the initial reaction rules is as follows:
\begin{lstlisting}{Join}
def x$_{\pt_1}$($\ldots$) & y$_1$($\ldots$) |> $P_1$
 or x$_{\pt_2}$($\ldots$) & y$_1$($\ldots$) |> $P_1$
 or x$_{\pt_2}$($\ldots$) & y$_2$($\ldots$) |> $P_2$
\end{lstlisting}
Observe that nondeterminism is now more explicit: an instance of
$\pt_2$ sent on channel~\lst"x" can be consumed by either the second
or the third reaction rule to trigger either $P_1$ or $P_2$.  We can
shorten the new definition a little by using \lst"or" in
join~patterns:
\begin{lstlisting}{Join}
def (x$_{\pt_1}$($\ldots$) or x$_{\pt_2}$($\ldots$)) & y$_1$($\ldots$) |> $P_1$
 or x$_{\pt_2}$($\ldots$) & y$_2$($\ldots$) |> $P_2$
\end{lstlisting}
Here the disjunctive composition ($\dis{J_1}{J_2}$) in join patterns
works as syntactic sugar, in the following sense: $$
\reaction{\para{J}{(\dis{J_1}{J_2})}}{P} \defineas
\dis{(\reaction{\para{J}{J_1}}{P})}{(\reaction{\para{J}{J_2}}{P})} $$

\item If $\pt_1 \equiv \pt_2$, then matching by their representative
is enough:
\begin{lstlisting}{Join}
def x$_{\repr{\pt_1}{\pt_2}}$($\ldots$) & y$_1$($\ldots$) |> $P_1$
 or x$_{\repr{\pt_1}{\pt_2}}$($\ldots$) & y$_2$($\ldots$) |> $P_2$
 or x(z) |> match z with
            | $\repr{\pt_1}{\pt_2}$ ->  x$_{\repr{\pt_1}{\pt_2}}$($\ldots$)
            | _ -> $\nullp$
\end{lstlisting}

\item Finally, if neither $\pt_1 \preceq \pt_2$ nor $\pt_2 \preceq
  \pt_1$ holds, with $\pt_1$ and $\pt_2$ being nevertheless compatible,
  then an extra matching by pattern $\lub{\pt_1}{\pt_2}$ is needed:
\begin{lstlisting}{Join}
def (x$_{\pt_1}$($\ldots$) or x$_{\lub{\pt_1}{\pt_2}}$($\ldots$)) & y$_1$($\ldots$) |> $P_1$
 or (x$_{\pt_2}$($\ldots$) or x$_{\lub{\pt_1}{\pt_2}}$($\ldots$)) & y$_2$($\ldots$) |> $P_2$
 or x(z) |> match z with
           | $\lub{\pt_1}{\pt_2}$ -> x$_{\lub{\pt_1}{\pt_2}}$($\ldots$)
           | $\pt_1$ -> x$_{\pt_1}$($\ldots$) | $\pt_2$ -> x$_{\pt_2}$($\ldots$)
           | _ -> $\nullp$
\end{lstlisting}
Note that the relative order of $\pt_1$ and $\pt_2$ is irrelevant here.
\end{enumerate}

In the transformation rules above, we paid little attention to
variables in patterns, by writing
\lst"x"$_{\pt}$\lst"("$\ldots$\lst")".  We now show variable
management by means of the concurrent stack example.  Here, the
relevant patterns are $\pt_1 = \id{ls}$ and $\pt_2 =
\cons{\id{x}}{\id{xs}}$ and we are in the case where $\pt_1 \preceq
\pt_2$ (and $\pt_2 \not\preceq \pt_1$ because of instance empty
list~$\nil$).  Our idea is to let dispatching focus on instance
checking, and to perform variable binding after synchronization:
\begin{lstlisting}{Join}
def pop(r) & State$_{\cons{\id{x}}{\id{xs}}}$(z) |> match z with x::xs -> r(x) & State(xs)
 or push(v) & (State$_{\cons{\id{x}}{\id{xs}}}$(z) or State$_{\id{ls}}$(z)) |> match z with ls -> State(v::ls)
 or State(z) |> match z with
                | _::_ -> State$_{\cons{\id{x}}{\id{xs}}}$(z)
                | _ -> State$_{\id{ls}}$(z)
\end{lstlisting}
One may believe that the matching of the pattern \lst"x::xs" needs to
be performed twice (once in the dispatcher, once in the first reaction
rule), but it is not necessary.  The compiler should know that the
matching of \lst$z$ against \lst$x::xs$ in the first reaction rule
cannot~fail, and as a consequence, no test needs to be performed here,
only the binding of the pattern variables.
See Section~\ref{subsec.optimization} for details.

 \section{The compilation \texorpdfstring{$\C{\cdot}$}{scheme}}
\label{sec.trans}

We formalize the intuitive idea described in Section~\ref{sec.trans-idea} as a
transformer $Y_x$, which transforms a join definition $D$ with respect to
channel $x$. The algorithm essentially works by constructing the meet
semi-lattice of the formal pattern arguments of channel~$x$ in $D$, modulo
pattern equivalence $\equiv$, with the less precise relation $\preceq$ being
the partial order.  Moreover, we visualize the lattice as a Directed Acyclic
Graph (DAG), namely, vertices as patterns, and edges representing the partial
order.  If we reason more on instance sets than on patterns, this structure is
quite close to the ``subset graph'' of~\cite{pritchard99computing}.
 
\medskip\noindent \textbf{Algorithm $Y_x$:}
Given $D$, a join definition, where $x$ is a channel defined by~$D$.
\begin{enumerate}[\ ]
\item {\bf Step 0: Preprocess.} \hspace*{1cm}
  \begin{enumerate}[(1)]
  \item Collect all the pattern arguments of channel $x$ into the
    sequence: $$\pts_x = \pt^{x}_1; \pt^x_2; \ldots; \pt^x_n$$
  \item Let $\pts'_x$ be formed from $\pts_x$ by replacing all
    variables by wildcards~``$\_$'' and taking the $\repr{}{}$ of all
    equivalent patterns; thus $\pts'_x$ is a sequence of pairwise
    nonequivalent patterns.
  \item Perform exhaustiveness check on $\pts'_x$, if not exhaustive,
    issue a warning.
  \item
    \begin{enumerate}[\ ]
    \item {\bf IF:}\ There is only one pattern in $\pts'_x$, and that
      $\pts'_x$ is exhaustive
    \item {\bf THEN:}\ goto Step 5. (In that case, no dispatching is needed.)
    \end{enumerate}
  \end{enumerate}
\item {\bf Step 1: Closure by least upper bound.} \hspace*{1cm} \\
  For any pattern~$\pt$ and pattern sequence $\pts = \pt_1 ; \pt_2 ;
  \ldots{} ; \pt_n$, we define $\lub{\pt}{\pts}$ as the sequence
  $\lub{\pt}{\pt_{i_1}} ; \lub{\pt}{\pt_{i_2}} ; \ldots ;
  \lub{\pt}{\pt_{i_m}}$, where the $\pt_{i_k}$\,s are the patterns from
  $\pts$ that are compatible with $\pt$.

  We also define function~$F$, which takes a pattern sequence~$\pts$
  as argument and returns a pattern sequence.
  \begin{enumerate}[\ ]
  \item {\bf IF:}\ $\pts$ is empty
  \item {\bf THEN:}\ $F(\pts)=\pts$
  \item {\bf ELSE:}\ Decompose $\pts$ as $\pt;\pts'$ and state $F(\pts) =
    \pt; F(\pts'); \lub{\pt}{F(\pts')}$
  \end{enumerate}
  Compute the sequence $\ptsbis' = F(\pts'_x)$.  It is worth noticing
  that $\ptsbis'$ is the sequence of all valid patterns
  $(\pt^{x'}_{i_1}\lubop\ldots
  (\lub{\pt^{x'}_{i_{k-1}}}{\pt^{x'}_{i_k}}) \ldots)$, with $1 \leq
  i_1 < i_2 < \ldots < i_k \leq n$, and $1 \leq k \leq n$, where we
  decompose $\pts'_x$ as $\pt^{x'}_1 ; \pt^{x'}_2 ; \ldots ;
  \pt^{x'}_{n}$.
\item {\bf Step 2: Up to equivalence.}\ \hspace*{1cm} \\
  As in Step $0.2$, build $\ptsbis$ by taking the $\repr{}{}$ of all
  equivalent patterns in $\ptsbis'$.
\item {\bf Step 3: Build DAG:} \hspace*{1cm} \\
  Corresponding to the semi-lattice $(\ptsbis, \preceq)$, build a
  directed acyclic graph $G(\V,\E)$.
  \begin{enumerate}[(1)]
  \item $\V=\emptyset, \E=\emptyset$.
  \item For each pattern~$\ptbis$ in $\ptsbis$, add a new vertex $v$
    into $\V$ and annotate the vertex with~$\ptbis$.
  \item $\forall (v, v') \in \V \times \V, v \ne v'$, with annotations
    $\ptbis$ and $\ptbis'$ respectively, if $\ptbis \preceq \ptbis'$,
    then add an edge from $v'$ to~$v$ into~$\E$.
  \end{enumerate}
\item {\bf Step 4: Add dispatcher.} \hspace*{1cm} \\
  Following one topological order, the vertices of $G$ are indexed as
  $v_1, \ldots, v_m$, correspondingly with annotations $\ptbis_1,
  \ldots, \ptbis_m$. We extend the join definition $D$ with a
  dispatcher on channel $x$ of the form: $x(z)$ \lst"|> match" $z$
  \lst"with" $\Lambda$, where $z$ is a fresh variable and $\Lambda$ is
  built as follows:
  \begin{enumerate}[(1)]
  \item Let $j$ ranges over $\set{1, \ldots, m}$. Following the
    topological order above, for all vertices~$v_j$ in $\V$ append a
    rule ``$\mid \ptbis_j \rightarrow x_{\ptbis_j}(z)$'' to~$\Lambda$,
    where $x_{\ptbis_j}$ is a fresh channel name assigned to
    vertex~$v_j$ whose annotation is $\ptbis_j$. Such fresh channels are here
    for the purpose of carrying messages originally sent to $x$ then
    forwarded by the dispatcher, hence are also referred to as
    \emph{forwarding channels}.
  \item If $\pts_x$ is not exhaustive, then add a rule ``$\mid \_
    \rightarrow \nullp$'' at the end.
  \end{enumerate}
\item {\bf Step 5: Rewrite reaction rules.} \hspace*{1cm} \\
  For each reaction rule defining channel $x$ in $D$:
  $\reaction{\para{J_i}{x(\pt^x_i)}}{Q_i}$, we rewrite it according to
  the following policy. Let $Q'_i = \matchone{z_i}{\pt^x_i}{Q_i}$, where
  $z_i$ is a fresh variable.
  \begin{enumerate}[\ ]
  \item {\bf IF:} coming from Step 0
  \item {\bf THEN:} rewrite to
    $\reaction{\para{J_i}{x(z_i)}}{Q'_i}$
  \item {\bf ELSE:}\hspace*{1cm}
    \begin{enumerate}[(1)]
    \item Let $v_{j_{i}}$ be the unique vertex in $\V$, s.t.  its
      annotation ${\ptbis_{j_{i}}} \equiv \pt^x_i$.
    \item We collect all the predecessors of $v_{j_{i}}$ in $G$, and
      we record the indices of them, together with $j_i$, into a set
      that we note $\preds{\pt^x_i}$.
    \item Rewrite to $\reaction{\para{J_i}
        {({\displaystyle\bigvee}_{j\in\preds{\pt^x_i}}x_{\ptbis_j}(z_i))}} {Q'_i}$,
      where $\displaystyle\bigvee$ is the generalized $\kwd{or}$ construct of
      join patterns.
    \end{enumerate}
    \end{enumerate}
    \end{enumerate}

Given a join definition~$D$, we note $\dv{D} = \set{x_1, \ldots, x_n}$ $(n
\geq 0)$, that is we order the channel names arbitrarily.
To transform $D$, we apply $Y_{x_n} \ldots Y_{x_1}(D)$. And the compilation of
processes $\C{\cdot}$ is inductively defined as follows: $$
\begin{array}{rcl}
\C{\nullp} & \defineas & \nullp \\
\C{x(\exp)} & \defineas & x(\exp) \\
\C{\para{P_1}{P_2}} & \defineas & \para{\C{P_1}}{\C{P_2}}\\
\C{\define{D}{P}} & \defineas & \define{Y_{x_n}\ldots Y_{x_1}(\C{D})}{\C{P}} \\
\C{\matchthree{\exp}{\pt}{P}} & \defineas & \matchC{\exp}{\pt}{P}\\
\\
\C{\top} & \defineas & \top \\
\C{\reaction{J}{P}} & \defineas & \reaction{J}{\C{P}} \\
\C{\dis{D_1}{D_2}} & \defineas & \dis{\C{D_1}}{\C{D_2}}
\end{array}
$$
Observe that the compilation preserves the interface of join
definitions. Namely, it only affects the join definitions, never
suppressing a channel, while message sending remains the same.

\section{Example of compilation}
\label{sec.example}

Given the following join definition for an enriched integer stack:
\begin{lstlisting}{Join}
def push(v) & State(ls) |> State (v::ls) 
 or pop(r) & State(x::xs) |> r(x) & State(xs)
 or insert(n) & State(0::xs) |> State(0::n::xs)
 or last(r) & State([x]) |> r(x) & State([x])
 or swap() & State(x$_1$::x$_2$::xs) |> State(x$_2$::x$_1$::xs)
 or pause(r) & State([]) |> r()
 or resume(r) |> State([]) & r()
\end{lstlisting}
The \lst"insert" channel inserts an integer as the second topmost
element, but only when the topmost element is $0$. The \lst"last"
channel gives back the last element in the stack, keeping the stack
unchanged.  The \lst"swap" channel exchange the topmost two elements
in the stack.  The \lst"pause" channel temporarily freezes the stack
when it is empty, while the \lst"resume" channel brings the stack back
into work.  We now demonstrate our transformation with respect to channel
\lst"State".
\def\one{1}
\def\two{2}
\def\three{3}
\def\four{4}
\def\five{5}
\def\six{6}
\def\seven{7}
\def\eight{8}
\begin{enumerate}[\ ]
\item {\bf Step 0:}\ We collect the pattern arguments of channel \lst"State"
  into $\pts_{\id{State}}$: $$ \pts_{\id{State}} = \id{ls};\
  \cons{\id{x}}{\id{xs}};\ \cons{0}{\id{xs}};\ \single{\id{x}};\
  \cons{\id{x}_1}{\cons{\id{x}_2}{\id{xs}}};\ \nil;\ \nil $$ We drop
  the last equivalent $\nil$ pattern during the up to equivalence
  substep~0.2, and we get: $$\pts'_{\id{State}} = \id{ls};\
  \cons{\id{x}}{\id{xs}};\ \cons{0}{\id{xs}};\ \single{\id{x}};\
  \cons{\id{x}_1}{\cons{\id{x}_2}{\id{xs}}};\ \nil $$ Additionally,
  $\pts'_{\id{State}}$ is exhaustive (pattern $\id{ls}$ alone covers
  all possibilities). Note that in the demonstration of this example,
  we sometimes keep variable names in patterns for readers'
  convenience. They are not necessary and are actually all replaced
  by~``$\_$'' in the implementation.
\item {\bf Step 1,2:} $\ptsbis'$ extends $\pts'_{\id{State}}$ with all
  possible least upper bounds. Then we form $\ptsbis$ from $\ptsbis'$
  by taking the $\repr{}{}$ of all equivalent patterns.  $$
  \ptsbis =
  \id{ls};\ \cons{\id{x}}{\id{xs}};\ \cons{0}{\id{xs}};\
  \single{\id{x}};\ \cons{\id{x}_1}{\cons{\id{x}_2}{\id{xs}}};\ \nil;\
  \cons{0}{\cons{\id{x}_2}{\id{xs}}};\ \single{0} $$
  Note that the
  last two patterns are new, where: 
  $$
  \begin{array}{rcccl}
    \cons{0}{\cons{\id{x}_2}{\id{xs}}} &&=&&
    \lub{\cons{0}{\id{xs}}}{\cons{\id{x}_1}{\cons{\id{x}_2}{\id{xs}}}}\\
    \single{0} &&=&& \lub{\cons{0}{\id{xs}}}{\single{\id{x}}}
  \end{array}  
  $$
\item {\bf Step 3:}\ We build the semi-lattice $(\ptsbis,\preceq)$, see
Figure~\ref{fig.ptjoin.dag}.
\begin{figure}[ht]
\centering
\input{dag.fex}
\caption{The semi-lattice of patterns and the topological order}\label{fig.ptjoin.dag}
\end{figure}
\item {\bf Step 4:}\ One possible topological order of the vertices is also
  given at the right of Figure~\ref{fig.ptjoin.dag}. Following that
  order, we build the dispatcher on channel~\lst"State".
\begin{lstlisting}{Join}
  or State(z) |> match z with
                 | 0::$\_$::$\_$ -> State$_{\one}$(z)
                 | [0] -> State$_{\two}$(z)
                 | $\_$::$\_$::$\_$ -> State$_{\three}$(z)
                 | 0::$\_$ -> State$_{\four}$(z)
                 | [$\_$] -> State$_{\five}$(z)
                 | $\_$::$\_$ -> State$_{\six}$(z)
                 | [] -> State$_{\seven}$(z)
                 | $\_$ -> State$_{\eight}$(z)
\end{lstlisting}
  where $\id{State}_{\one}$, $\ldots$, $\id{State}_{\eight}$ are the
  fresh forwarding channels.
\item {\bf Step 5:}\ We rewrite the original reaction rules.  As an example,
  consider the third reaction rule for the \lst"insert" behavior: the
  pattern in \lst"State(0::xs)" corresponds to vertex $4$ with
  annotation $\cons{0}{\id{xs}}$ in the graph, which has two
  predecessors: vertex $1$ with annotation
  $\cons{0}{\cons{\id{x}_2}{\id{xs}}}$ and vertex $2$ with annotation
  $\single{0}$.  Therefore, the reaction rule is rewritten to:
\begin{lstlisting}{Join}
insert(n) & (State$_{\one}$(z$_3$) or State$_{\two}$(z$_3$) or State$_{\four}$(z$_3$)) 
            |> match z$_3$ with 0::xs -> State(0::n::xs)
\end{lstlisting}
where $\id{z}_3$ is a fresh variable.
\end{enumerate}
As a final result of our transformation, we get the disjunction of the
following rules and of the dispatcher built in Step 4.

\begin{lstlisting}{Join}
def push(v) & (State$_{\one}$(z$_1$) or $\dots$ or State$_{\eight}$(z$_1$)) 
              |> match z$_1$ with ls -> State (v::ls) 
 or pop(r) & (State$_{\one}$(z$_2$) or $\ldots$ or State$_{\six}$(z$_2$))
              |> match z$_2$ with x::xs -> r(x) & State(xs)
 or insert(n) & (State$_{\one}$(z$_3$) or State$_{\two}$(z$_3$) or State$_{\four}$(z$_3$))
              |> match z$_3$ with 0::xs -> State(0::n::xs)
 or last(r) & (State$_{\two}$(z$_4$) or State$_{\five}$(z$_4$))
              |> match z$_4$ with [x] -> r(x) & State([x])
 or swap() & (State$_{\one}$(z$_5$) or State$_{\three}$(z$_5$))
              |> match z$_5$ with x$_1$::x$_2$::xs -> State(x$_2$::x$_1$::xs)
 or pause(r) & State$_{\seven}$(z$_6$) |> match z$_6$ with [] -> r()
 or resume(r) |> State([]) & r()
\end{lstlisting}

\section{Correctness}
\label{sec.correct}

A program written in the applied join calculus of
Section~\ref{sec.joinpi} is a process $P$. The compilation $\C{P}$
replaces all the join definitions $D$ in $P$ by $Y_{x_n}\ldots
Y_{x_1}(D)$, where $\dv{D} = \set{x_1,\ldots,x_n}$. To guarantee the
correctness, we require the programs before and after the compilation
be open equivalent. Namely, the following theorem should hold.
\begin{thm}\label{th.joinpi.correct}
  For any process $P$, $\C{P} \pseq P$.
\end{thm}
\begin{proof}
  By structural induction on processes. Because $\pseq$ is a full
  congruence and a transitive relation, it suffices to prove one step
  of the compilation, that is, $Y_x$ is correct (see
  Lemma~\ref{lemma.Y} below). \forget{\qed}
\end{proof}
\begin{lem}\label{lemma.Y}
  For any join definition $D$, channel name $x \in \dv{D}$, and
  process $P$, we have:
  $$\define{D}{P} \pseq \define{Y_x(D)}{P}$$
\end{lem}
This lemma is crucial to the correctness of the compilation. We elaborate the proof in the coming sections.
First, we recall the notations of algorithm $Y_x$ in
Section~\ref{subsec.Yx}. Then, we discuss the properties of the
dispatcher built by $Y_x$ in Section~\ref{subsec.dispatcher}. Finally,
we prove Lemma~\ref{lemma.Y} in Section~\ref{subsec.prooflemmaY}.

\subsection{Summary of notations}
\label{subsec.Yx}

We summarize the connection between the input and the output of $Y_x$.
For simplicity, we omit the $x$ superscripts everywhere.
According to the algorithm given in Section~\ref{sec.trans},
there are two cases during the procedure of $Y_x$, chosen at the end
of Step 0:
\paragraph{\bf Case ``jump''}\ The case where Steps 1 to 4 are skipped. Then,
  for any reaction rule of the form $\reaction{\para{J_i}{x(\pt_i)}}{Q_i}$
  of $D$, $i=1\ldots n$, the pattern $\pt_i$ is irrefutable, namely,
  $\pt_i \equiv \_$. And in $Y_x(D)$, we have the corresponding
  reaction rule
  $\reaction{\para{J_i}{x(z_i)}}{\matchone{z_i}{\pt_i}{Q_i}}$, where $z_i$
  is fresh.
\paragraph{\bf Case ``go through''}\ The general case. We recall the notations
  of the DAG $G(\V,\E)$ built by the algorithm. $G$ has $m$ vertices,
  and following the topological order, the vertices are indexed as
  $v_1, \ldots, v_m$ with pattern annotations $\ptbis_1, \ldots,
  \ptbis_m$.  Each vertex is
  also assigned a fresh forwarding channel, written $x_{\ptbis_j}$.
  
  For any reaction rule of the form
  $\reaction{\para{J_i}{x(\pt_i)}}{Q_i}$ of $D$, $i=1\ldots n$, there
  exists a unique vertex in $G$ called $v_{j_{i}}$, such that its
  annotation $\ptbis_{j_{i}} \equiv \pt_i$. We use $\preds{\pt_i}$
  to record the indices of the predecessors of $v_{j_{i}}$ as well as
  $j_i$. Note that we have $\pt_i \preceq \ptbis_j$ iff
  $j\in\preds{\pt_i}$. In $Y_x(D)$, we have a corresponding reaction
  rule as $
  \reaction{\para{J_i}{(\displaystyle\bigvee_{j\in\preds{\pt_i}}x_{\ptbis_j}(z_i))}}
  {\matchone{z_i}{\pt_i}{Q_i}} $, where the variable $z_i$ is fresh.
  Moreover, we add a dispatcher on channel $x$ into $Y_x(D)$ as:
\begin{lstlisting}{Join}
x(z) |> match z with
        | $\ptbis_1$ -> $x_{\ptbis_1}(z)$
        | $\ldots$
        | $\ptbis_m$ -> $x_{\ptbis_m}(z)$
        | _ -> $\nullp$              (* if non-exhaustive *) 
\end{lstlisting}
where $z$ is a fresh variable.

\subsection{Property of the dispatcher}
\label{subsec.dispatcher}

We go on to discuss the property of the dispatcher built during the
transformation on channel~$x$. Let $u$ range over closed expressions,
that is over values. Modulo pattern equivalence $\equiv$, the patterns
of the dispatcher ($\ptbis_j, j=1,\ldots,m$) are all the least upper
bounds of the pattern arguments of channel $x$ in the original~$D$
($\pt_i, i=1,\ldots,n$). Thus, the $\pt_i$\,s and the $\ptbis_j$\,s
admit the same instances: $\bigcup_{1 \leq i \leq n} \ins{\pt_i} =
\bigcup_{1 \leq j \leq m} \ins{\ptbis_j}$.  As an immediate
consequence, on one hand, for the set of values that do not match any
of the original~$\pt_i$\,s, written $\aleph = \{ u \mid \forall i, u
\not\in \ins{\pt_i}\}$, the values of $\aleph$ do not match
any $\ptbis_j$ either, and those values are silently eaten by the
dispatcher. On the other
hand, given any value~$u$ such that there exists at least one $\pt_i$
with $u \in \ins{\pt_i}$, then the dispatcher must forward $u$ onto
one of the forwarding channels. More precisely, the following lemma holds.
\begin{lem}\label{lemma.joinpi.dispatcher}
  For any value~$u$ that is an instance of some original pattern
  argument~$\pt_i$, the dispatcher forwards~$u$ to the forwarding
  channel assigned to a vertex in $G$, whose index belongs to
  $\preds{\pt_i}$.
\end{lem}
\begin{proof}
  We thus assume $u \in \ins{\pt_i}$.  Let $K$ be the set of indices $\{ k
  \mid u \in \ins{\pt_k} \}$ and $\pts_{K} = \set{\pt_k \mid k\in K}$. Let
  $\ptbis$ be the least upper bound of the patterns in
  $\pts_{K}$, written $\lubop{\pts_K}$ ($\ptbis$ exists, since
  $\pts_{K}$ is non-empty).  By steps 1--3 of
  the compilation algorithm~$Y_x$, there must exist some vertex denoted by
  $v_{j_{K}}$ in $G$ with annotation $\ptbis_{j_{K}} \equiv \ptbis$.    
  The dispatcher forwards message~$u$ onto the forwarding
  channel~$x_{\ptbis_{j_{K}}}$, for the following two
  reasons.
  \begin{enumerate}[(1)]
  \item Value~$u$ is an instance of $\ptbis_{j_{K}}$.
  \item No pattern of the dispatcher that appears before
  $\ptbis_{j_{K}}$ admits~$u$ as an instance.
  Namely, any pattern of the dispatcher $\ptbis_{j}, 1\leq j \leq m$,
  such that $u\in\ins{\ptbis_j}$ must be the least upper
  bound of a subset of~$\pts_K$. Then, since the
  patterns of the dispatchers are ordered topologically (with
  precision order $\preceq$), $\ptbis_{j_K}$ must be the foremost
  pattern in the dispatcher which has $u$ as an instance.
  Namely, precision order $\preceq$ applied to least upper bounds is
  reverse set inclusion applied to instance sets.
  \end{enumerate}
  Moreover, because $\pt_i\in\pts_K$ and $\ptbis_{j_K} \equiv \ptbis =
  \lubop\pts_K$, we have $\pt_i \preceq \ptbis_{j_{K}}$.
  Thus, by definition of~$I(\pt_i)$, we have $j_{K} \in I(\pt_i)$. \forget{\qed}
\end{proof}

In the following, given some value~$u$, we write $x_u$ for the forwarding
channel to which $u$ is sent by the dispatcher.  Using the new notation,
Lemma~\ref{lemma.joinpi.dispatcher} is reformulated as follows: if $u \in
\ins{\pt_i}$, then $x_u$ exists and we have $x_u\in\set{x_{\ptbis_j} \mid
  j\in\preds{\pt_i}}$.

\subsection{Proof of Lemma~\ref{lemma.Y}}
\label{subsec.prooflemmaY}
\forget{Now we are ready to prove Lemma~\ref{lemma.Y}.}
\resetequationcounter
\begin{proof}
  Following the
  definition of $\pseq$, we should prove ${(\define{D}{P})}{\sigma} \wbc
  {(\define{Y_x(D)}{P})}{\sigma}$, for any closing substitution~$\sigma$.
  In other words, since ${Y_x(D)}{\sigma} = Y_x({D}{\sigma})$,
  we should prove:
  \begin{align}
    \define{{D}{\sigma}}{{P}{\sigma_1}} \wbc
    \define{Y_x({D}{\sigma})}{{P}{\sigma_1}}
    \label{dagger}
  \end{align}
  where $\sigma_1$ is $\sigma$ minus the (possible) bindings of the
  variables of $\dv{D}$.
  Notice that all subcomponents ${D}{\sigma}$, $Y_x({D}{\sigma})$ and
  ${P}{\sigma_1}$ are closed. 
  Hence, to prove
  \eqref{dagger}, it suffices to prove that $Y_x$ is correct for
  closed terms (Lemma~\ref{lemma.joinpi.Y.close} below). \forget{\qed}
\end{proof}

\begin{lem}\label{lemma.joinpi.Y.close}
  For any closed join definition $D$, channel name $x\in\dv{D}$, and
  closed process $P$, we have:
  $$
  \define{D}{P} \wbc \define{Y_x(D)}{P}
  $$
\end{lem}
\resetequationcounter
\begin{proof}
  There are two subcases.
  \paragraph{\bf Case ``go through''}
  We construct the following relation $\R$: $$ \R =
  \set{(\define{D}{(\para{P}{Q})},\;
  \define{Y_x(D)}{(\para{P}{\bisi{Q}})})} $$ Above, process~$P$ and
  definition~$D$ range respectively over closed processes and closed
  definitions; while $Q$ and $\bisi{Q}$ are particular.  Dissect the
  structure of $D$ as: $$D =
  \dis{\ldots}{\dis{\reaction{\para{J_i}{x(\pt_i)}}{Q_i}}{\ldots}}$$
  We define $Q$ and $\bisi{Q}$ to be:
  \begin{align*}
    Q &= ({\prod}_{\delta \in \Delta}x({\pt_i}{\delta})) \; \& \;
    ({\prod}_{\psi \in \Psi}{Q_i}{\psi}) \;\&\; ({\prod}_{u \in
      U}x(u))\\
    \bisi{Q} &= ({\prod}_{\delta \in
      \Delta}x_{{{\pt_i}{\delta}}}({\pt_i}{\delta})) \; \& \;
    ({\prod}_{\psi \in \Psi}
    \matchone{{\pt_i}{\psi_{\pt_i}}}{\pt_i}{{Q_i}{\psi_{J_i}}})
  \end{align*}
  We note $\displaystyle\prod$ the generalized parallel composition.
  Note that processes $Q$ and  $\bisi{Q}$ are (implicitly) parameterized by
  the multisets of substitutions
  $\Delta$ and  $\Psi$, and by the multiset of values~$U$.
  In the definition of~$\R$, $\Delta$, $\Psi$ and $U$ range over all
  appropriate multisets.
  More precisely, given any reaction rule
  $\reaction{\para{J_i}{x(\pt_i)}}{Q_i}$ from $D$,
  we note $\delta$ any (closed) substitution on domain $\rv{\pt_i}$.
  Then, $\Delta$ stands for any multiset of such substitutions~$\delta$.
  Similarly, let $\psi$ be a (closed) substitution on domain
  $\rv{J_i}\uplus\rv{\pt_i}$.
  Moreover, for any such~$\psi$, let
  $\psi_{\pt_i}$ be ${\psi}\restrict{\rv{\pt_i}}$ (the restriction of
  $\psi$ on domain $\rv{\pt_i}$), and $\psi_{J_i}$ be
  ${\psi}\restrict{\rv{J_i}}$. Because $\rv{\pt_i}\cap\rv{J_i}=\emptyset$,
  the substitution $\psi$ is the sum of $\psi_{\pt_i}$ and $\psi_{J_i}$,
  written $\psi=\psi_{\pt_i} \uplus \psi_{J_i}$, and we further
  require  $\composite{\psi_{\pt_i}}{\psi_{J_i}} = \psi_{J_i} \uplus
  \psi_{\pt_i}$.
  Then, $\Psi$ is any multiset of such substitutions~$\psi$.
  Finally, $U$ is a multiset of elements from~$\aleph$.
    \begin{figure}
      \centering
        \input{bisi.fex}
      \caption{Reduction chasing in case ``go through''}\label{fig.joinpi.bisi}
    \end{figure}

  Intuitively, we use $Q$ and $\bisi{Q}$ to bridge the differences
  caused by $D$ and $Y_x(D)$. More specifically: a message
  $x({\pt_i}{\delta})$ may be forwarded to
  $x_{{{\pt_i}{\delta}}}({\pt_i}{\delta})$ by the dispatcher in
  $Y_x(D)$; furthermore, if a guarded process ${Q_i}{\psi}$ is
  triggered from $D$, then from $Y_x(D)$, we have the corresponding
  guarded process
  $\matchone{{\pt_i}{\psi_{\pt_i}}}{\pt_i}{{Q_i}{\psi_{J_i}}}$
  triggered; finally, a message on channel $x$ with a non-matching
  content, that is from $\aleph$, will be eaten by $Y_x(D)$.

  We analyze the following three aspects of $\R$: closure by (closed)
  evaluation contexts; reduction bisimulation; and preservation of barbs.

  \begin{enumerate}[$\bullet$]
  \item $\R$ is closed by closed evaluation contexts up to $\equiv$
    (\ie Lemma~\ref{lemma.wbc-upto-equiv-alt}.(1)). For any closed
    evaluation context $\econtext{\cdot}$, we have:
    \begin{align*}
      \econtext{\define{D}{(\para{P}{Q})}} &\equiv 
      \define{\dis{D}{D'}}{(\para{\para{P}{P'}}{Q})} \\
      \econtext{\define{Y_x(D)}{(\para{P}{\bisi{Q}})}} &\equiv
      \define{\dis{Y_x(D)}{D'}}{(\para{\para{P}{P'}}{\bisi{Q}})}
    \end{align*}
    where $\dv{D} \cap \dv{D'} = \emptyset$, so that $\dis{Y_x(D)}{D'} =
    Y_x(\dis{D}{D'})$. Therefore, we have
    $\econtext{\define{D}{(\para{P}{Q})}} \equiv\R\equiv
    \econtext{\define{Y_x(D)}{(\para{P}{\bisi{Q}})}} $.
  \item $\R$ is a reduction bisimulation (\ie a special case of
    Lemma~\ref{lemma.wbc-upto-equiv-alt}.(2) because the identity in
    included in~$\equiv$). We only detail the nontrivial cases.
    \begin{enumerate}[(1)]
    \item If there is a message $x({\pt_i}{\delta'})$ in $P$, the
      right part can forward it to a message
      $x_{{{\pt_i}{\delta'}}}({\pt_i}{\delta'})$ by the dispatcher
      in $Y_x(D)$. This reduction is simulated in the left part by no
      reduction, and we add the new substitution $\delta'$ into
      $\Delta$.
    \item\label{2} Similarly, if there is a message $x(u')$ in $P$,
      for some $u' \in \aleph$, the right part can eat the message by
      the dispatcher in $Y_x(D)$. This reduction is simulated by no
      reduction in the left part and we add $u'$ into $U$.
    \item If a reduction according to the reaction rule
      $\reaction{\para{J_i}{x(\pt_i)}}{Q_i}$ consumes a molecule
      $\para{{J_i}{\eta}}{x({\pt_i}{\delta})}$ in the left part, for
      some $\delta \in \Delta$ (\ie $x({\pt_i}{\delta})$
      occurs in $Q$) and $J_i\eta$ from $P$, with
      $\dom{\eta}=\rv{J_i}$; it can be simulated by consuming
      $\para{{J_i}{\eta}}{x_{{{\pt_i}{\delta}}}({\pt_i}{\delta})}$ in
      the right part, using the corresponding reaction rule
      $\reaction{\para{J_i}{(\displaystyle\bigvee_{j\in
            \preds{\pt_i}}x_{\ptbis_j}(z_i))}}
      {\matchone{z_i}{\pt_i}{Q_i}}$, because $x_{\pt_i\delta} \in
      \set{x_{\ptbis_j}\mid j\in\preds{\pt_i}}$
      (Lemma~\ref{lemma.joinpi.dispatcher}). The derivatives are still
      in $\R$, with $\Delta$ shrinking to $\Delta\setminus\set{\delta}$,
      and $\Psi$ expanding to $\Psi\cup\set{\eta\uplus\delta}$. We
      assume $\alpha$-conversion when necessary to guarantee
      $\composite{\delta}{\eta}={\eta\uplus\delta}$.  Vice versa.
    \item Similar to the previous case but this time the left part
      consumes a molecule $\para{{J_i}{\eta}}{x({\pt_i}{\delta'})}$,
      where $\delta'$ is not from $\Delta$. Then, the right part
      simulates this reduction by first forwarding the message
      $x({\pt_i}{\delta'})$ to the message
      $x_{{{\pt_i}{\delta'}}}({\pt_i}{\delta'})$ as in case~\ref{2},
      then consuming the molecule
      $\para{{J_i}{\eta}}{x_{{{\pt_i}{\delta'}}}({\pt_i}{\delta'})}$.
      $\Psi$ expands to $\Psi\cup\set{\eta\uplus\delta'}$.
    \item The
      $\matchone{{\pt_i}{\psi_{\pt_i}}}{\pt_i}{{Q_i}{\psi_{J_i}}}$
      in $\bisi{Q}$ of the right part can be reduced to
      ${({Q_i}{\psi_{J_i}})}{\psi_{\pt_i}}$ by the semantic rule
      \rln{Match}.  Because we have
      $\composite{\psi_{\pt_i}}{\psi_{J_i}}=\psi_{J_i}\uplus\psi_{\pt_i}$,
      the result of the reduction equals to
      ${Q_i}{(\psi_{J_i}\uplus\psi_{\pt_i})}$, that is
      ${Q_i}{\psi}$. This reduction is simulated by no reduction in
      the left part.  However, the process $P$ becomes
      $\para{P}{{Q_i}{\psi}}$, and $\Psi$ shrinks to
      $\Psi\setminus\set{\psi}$.
    \item If a reduction involves ${Q_i}{\psi}$ from $Q$ of the left
      part, for some $\psi\in \Psi$, it can be simulated by first
      reducing the correspondent
      $\matchone{{\pt_i}{\psi_{\pt_i}}}{\pt_i}{{Q_i}{\psi_{J_i}}}$
      from $\bisi{Q}$ into ${Q_i}{\psi}$ as in the previous case.
    \end{enumerate}
      Figure~\ref{fig.joinpi.bisi} summarizes the various cases we
      just examined, where thick lines
      express the $\R$~relation.
    \item $\R$ preserves barbs (\ie
      Lemma~\ref{lemma.wbc-upto-equiv-alt}.(3)). We demonstrate
      $\barb{\define{D}{(\para{P}{Q})}}{y} \implies
      \weakbarb{\define{Y_x(D)}{(\para{P}{\bisi{Q}})}}{y}$ and vice
      versa. We distinguish the cases that make
      $\barb{\define{D}{(\para{P}{Q})}}{y}$ hold.
    \begin{enumerate}[(1)]
    \item $\barb{Q}{y}$. We have $y\not\in\dv{D}$. Because all
      variables in $\dv{Y_x(D)}\setminus\dv{D}$ are fresh, we also
      have $y\not\in\dv{Y_x(D)}$. According to the structure of $Q$,
      we must have $\barb{Q_i\psi}{y}$ for some $\psi\in\Psi$.
      Then in $\bisi{Q}$, we have $
      \matchone{{\pt_i}{\psi_{\pt_i}}}{\pt_i}{{Q_i}{\psi_{J_i}}}$
      reduces to $Q_i\psi$ and $\barb{Q_i\psi}{y}$. That is,
      $\weakbarb{(\matchone{{\pt_i}{\psi_{\pt_i}}}{\pt_i}{{Q_i}{\psi_{J_i}}})}{y}$,
      \ie $\weakbarb{\bisi{Q}}{y}$, \ie
      $\weakbarb{\define{Y_x(D)}{(\para{P}{\bisi{Q}})}}{y}$.
    \item $\barb{P}{y}$. Obvious.
    \end{enumerate}
    The proof of the other direction, \ie
    $\barb{\define{Y_x(D)}{(\para{P}{\bisi{Q}})}}{y} \implies
    \weakbarb{\define{D}{(\para{P}{Q})}}{y}$, is obvious since the
    only case for
    $\barb{\define{Y_x(D)}{(\para{P}{\bisi{Q}})}}{y}$ is when
    $\barb{P}{y}$.

    Following the analysis above, $\R$ is a weak barbed congruence up
    to $\equiv$. By Lemma~\ref{lemma.wbc-upto-equiv}, we have $\R$ is a
    weak barbed congruence.

    Let $\Delta$, $\Psi$ and $U$ be empty sets. We have the two
    processes of Lemma~\ref{lemma.joinpi.Y.close} satisfy relation
    $\R$, hence $\wbc$. That is, we proved that
    Lemma~\ref{lemma.joinpi.Y.close} holds for case ``go through''.
  \end{enumerate}
    \begin{figure}
      \centering
        \input{bisi2.fex}
        \caption{Reduction chasing in case
          ``jump''}\label{fig.joinpi.bisi2}
    \end{figure}
  \paragraph{\bf Case ``jump''} 
  We build another relation $\R$, with $Q$ and $\bisi{Q}$ defined as
  follows:
  \begin{align*}
    Q &= {\prod}_{\psi \in \Psi}{Q_i}{\psi}\\
    \bisi{Q} &= {\prod}_{\psi \in \Psi}
    \matchone{{\pt_i}{\psi_{\pt_i}}}{\pt_i}{{Q_i}{\psi_{J_i}}}
  \end{align*}
  and we summarize the property of reduction bisimulation by the diagram of
  Figure~\ref{fig.joinpi.bisi2}. \forget{\qed}
\end{proof}

\section{Implementing applied join}
\label{sec.imp}

We carried out the practical implementation work of the applied join
calculus as an extension of the \jocaml system.
The extended system is publicly released~\cite{JoCaml}.
The release includes a tutorial that makes extensive use of
algebraic patterns in join patterns.
In this section, we
first sketch out the structure of the extended \jocaml compiler,
pointing out where the transformation should take place. Then some
optimizations of our algorithm $Y_x$ are reported.

\subsection{Front end of the (extended) \jocaml compiler}
\label{subsec.frontend}

The \jocaml compiler is an extension of the \ocaml compiler,
as the \jocaml language is an extension of the \ocaml language.
Extensions are confined to the first four phases of the compiler.

More precisely, there are additional tokens in the lexer (such as the
keyword \lst|def|).  Then, all the constructs of
Figure~\ref{fig.syntax} are parsed and rendered as specific constructs
in the abstract syntax tree. Typed syntax undergoes a
similar extension. Amongst those first three compiler phases, only the
typer significantly differs from the original \ocaml compiler, since
the \jocaml compiler has to deal with the specific rules for typing
the join calculus
polymorphically~\cite{Fournet-Laneve-Maranget-Remy:typing-join}.
Finally, the typed syntax is translated to \emph{lambda-code}, which
basically is $\lambda$-calculus enriched with primitive types and calls
to the runtime library.  All constructs specific to \jocaml disappear,
being replaced by calls to specific primitives in a ``\textsf{Join}''
library, built on top of one of the \ocaml thread libraries.  In the
following, we denote as ``the \jocaml runtime'', the ordinary (thread
aware) \ocaml runtime, plus the thread library, plus the \textsf{Join}
library. To summarize, extending the \ocaml system to the \jocaml
system amounts to modifying the front end of the compiler, and to
writing the \textsf{Join} library.

\begin{figure}[ht]
\centering
\input{ejocaml_flow.fex}
\caption{The extended \jocaml compiler front end}\label{fig.ejocaml.flow}
\end{figure}
Extending \jocaml to handle pattern arguments in join definitions
requires further modifications.  Figure~\ref{fig.ejocaml.flow} shows
the structure of the extended \jocaml compiler. With respect to plain
\jocaml (without algebraic pattern matching in join definitions), the
parser and the typer have to be modified to take pattern arguments in
channel definitions into account. However these extensions are
mechanical. The critical modification manifests itself as an extra
sub-phase (enclosed in the dashed polygon) between the typing phase and
the translation phase. Not surprisingly, the additional phase carries
out the transformation from extended join definitions to plain ones,
by implementing the compilation scheme $\C{\cdot}$ of
Section~\ref{sec.trans}.  Once this new transformation is performed,
all join definitions in the typed trees are plain ones (without
pattern arguments).  Then, the translator to lambda-code and, more
importantly, the \jocaml runtime system need not be changed, with
respect to the ones of the original \jocaml system.

We in fact also slightly extended the translator, for the sake of
performing a few optimizations (see Section~\ref{subsec.optimization})
and of avoiding excessive duplications of guarded processes (see
Section~\ref{subsec.share}). The optimizations we perform make use of
the sophisticated pattern matching compiler and analyzer that are
already present in the standard \ocaml compiler.

\subsection{Matching optimizations}
\label{subsec.optimization}

\subsubsection{Avoiding redundant matchings}
As discussed at the end of Section~\ref{sec.trans-idea},
the compilation introduces redundant matchings.
For instance, in the stack example, we get:
\begin{lstlisting}{Join}
def pop(r) & State$_{\cons{\id{x}}{\id{xs}}}$(z) |> match z with x::xs -> r(x) & State(xs)
  $\ldots$
or State(y) |> match y with
               | _::_ -> State$_{\cons{\id{x}}{\id{xs}}}$(y)
                         $\ldots$
\end{lstlisting}
A $\id{pop}$ operation apparently involves matching the $\id{State}$
argument twice: once in the dispatcher, to select the appropriate
forwarding channel~\lst|State$_{\cons{\id{x}}{\id{xs}}}$|, and again
in the reaction rule, to perform the bindings of variables \lst|x| and
\lst|xs| to the head and tail of the cons-cell~\lst|z|.

However, by construction, the value of argument~\lst|z| is guaranteed
to be an instance of the pattern \lst|x::xs|. This remark is general
(see Lemma~\ref{lemma.joinpi.dispatcher}): for any matching
$\matchone{z_i}{\pt_i}{Q_i}$ introduced in reaction rules by Step~5 of
algorithm~$Y_x$, the value of $z_i$ always matches the
pattern~$\pt_i$.  In other words, the matching
$\matchone{z_i}{\pt_i}{Q_i}$ never fails, hence no test need to be
performed at all. As a consequence, in the case of the pattern
\lst|x::xs|, we aim at getting the the following
lambda-code:\footnote{In examples, we show lambda-code as \ocaml code,
enriched with a few primitives.}
\begin{lstlisting}{Join}
let x = field 0 z in 
let xs = field 1 z in
$\ldots$
\end{lstlisting}
Primitives ``$\id{field}\; 0 \; \id{z}$'' and ``$\id{field}\; 1 \;
\id{z}$'' extract the head and tail from the cons-cell~$\id{z}$.

The requirement is then to write a specific matching compiler that
does not issue tests when test outputs can be predicted at compile
time. In fact, such a matching compiler is already present in the
\ocaml compiler: as it stands, the optimizing pattern matching
compiler of~\cite{LefessantMarangetPattern} can output such code,
provided it is informed that the compiled matching has only one clause
and never fails, which is exactly the case for all the matchings
$\matchone{z_i}{\pt_i}{Q_i}$ introduced in reaction rules by Step~5 of
algorithm~$Y_x$.
Incidentally, the condition ``the matching can never fail'' is
expressed simply as ``the matching is exhaustive''.
We also rely on a later phase of the \ocaml compiler
to inline \lst|let|-bindings when appropriate. 

As a final remark, it is worth observing that, when the original
pattern does not contain variables, the compilation of
$\matchone{z_i}{\pt_i}{\cdots}$ yields no code: neither test, nor
binding.

\subsubsection{Avoiding useless forwarding channels}
\label{subsubsec.remove}
Simple analysis of the dispatcher matching enables use to spare some
of the forwarding channels.  Let us first re-consider the example of
the complete stack. Our transformer $Y$ applied to channel~\lst|State|
yields the following dispatcher:
\begin{lstlisting}{Join}
or State(z) |> match z with
                 | 0::_::_ -> State$_{\one}$(z)
                 | [0] -> State$_{\two}$(z)
                 | _::_::_ -> State$_{\three}$(z)
                 | 0::_ -> State$_{\four}$(z)
                 | [_] -> State$_{\five}$(z)
                 | _::_ -> State$_{\six}$(z)
                 | [] -> State$_{\seven}$(z)
                 | _ -> State$_{\eight}$(z)
\end{lstlisting}
In the matching above, some clauses are never matched at runtime.  For
instance, the last clause ``\lst|_ -> State$_{\eight}$(z)|'' is
useless, because of the two immediately preceding clauses
``\lst"_::_->"~$\ldots$'' and ``\lst"[] ->"~$\ldots$'' that obviously
match all the lists.  As a consequence, the forwarding channel
$\id{State}_{\eight}$ never carries any message hence it is also
useless.  Similarly, channels $\id{State}_{\four}$ and channel
$\id{State}_{\six}$ are useless. We can optimize by removing both the
useless clauses from the dispatcher and all occurrences of useless
channels from the rewritten join patterns.

To summarize, by applying the optimizations discussed so far, the
stack example after compilation looks as follows:
\begin{lstlisting}{Join}
def push(v) & (State$_{\one}$(z$_1$) or State$_{\two}$(z$_1$) or State$_{\three}$(z$_1$) or State$_{\five}$(z$_1$) or State$_{\seven}$(z$_1$))
      |> State (v::z$_1$) 
 or pop(r) & (State$_{\one}$(z$_2$) or State$_{\two}$(z$_1$) or State$_{\three}$(z$_1$) or State$_{\five}$(z$_1$))
      |> r(field 0 z$_2$) & State(field 1 z$_2$)
 or insert(n) & (State$_{\one}$(z$_3$) or State$_{\two}$(z$_3$))
      |> State(0::n::field 1 z$_3$)
 or last(r) & (State$_{\two}$(z$_4$) or State$_{\five}$(z$_4$))
      |> let x = field 0 z$_4$ in r(x) & State([x])
 or swap() & (State$_{\one}$(z$_5$) or State$_{\three}$(z$_5$))
      |> let m = field 1 z$_5$ in State(field 0 m::field 0 z$_5$::field 1 m)
 or pause(r) & State$_{\seven}$(z$_6$) |> r()
 or resume(r) |> State([]) & r()
 or State(z) |> match z with
                 | 0::$\_$::$\_$ -> State$_{\one}$(z)
                 | [0] -> State$_{\two}$(z)
                 | $\_$::$\_$::$\_$ -> State$_{\three}$(z)
                 | [$\_$] -> State$_{\five}$(z)
                 | [] -> State$_{\seven}$(z)
\end{lstlisting}
Thanks to the optimization, three cases are spared from the
dispatcher, three channels are not allocated, and the size of the
$\kwd{or}$~join-patterns decrease significantly.

To integrate this optimization into the implementation, we modify the
algorithm $Y_x$, as regards dispatcher construction (Step 4) and
rewriting of reaction rules (Step 5). In Step 4, after the topological
sort, we check the usefulness of each vertex.  More specifically, to
check whether vertex $v_k$ is useful or not, with respect to the
preceding vertices $v_1$, \ldots{}, $v_{k-1}$ in the topological
order, we check the usefulness of pattern $\ptbis_k$ with respect to
patterns $\ptbis_1$, \ldots{}, $\ptbis_{k-1}$, where $\ptbis_i$ is the
annotation pattern of vertex~$v_i$. For that purpose, we use the
standard usefulness checker of \ocaml~\cite{warning}, of which we
present a simplified version in Section~\ref{subsec.pt}.  Then, in
Step~5 of the algorithm we retain only the $v_k$'s that are useful.

\subsection{Compiling \textbf{or} in join patterns}
\label{subsec.share}

The compilation scheme $\C{\cdot}$ introduces disjunctive composition
into join patterns, a construct that \jocaml did not support before
the introduction of pattern argument in join definitions. In this
section, we describe our extensions to the \jocaml compiler so as to
integrate this new feature.

When we introduced \kwd{or} in join patterns, we claimed that it is
syntactic sugar. That is, we define this new construct by distributing 
$\kwd{\&}$ over $\kwd{or}$, until \kwd{or} reaches the reaction rule
level, where we finally duplicate the reaction rules themselves.
$$
\reaction{(J_1 \infix{or} J_2 \infix{or} \cdots \infix{or} J_n)}{P}
\defineas
(\reaction{J_1}{P}) \infix{or} 
(\reaction{J_2}{P}) \infix{or} \cdots
 \infix{or} (\reaction{J_n}{P})
$$
The whole process of distributing $\kwd{\&}$ over $\kwd{or}$ and of
duplicating the rules can be summarized as ``expansion of \kwd{or} in
join patterns''.

It is not difficult to see that the above mentioned expansion easily
produces an exponential number of reaction rules.
For instance, consider the definition:
\begin{lstlisting}{Join}
def a$^1$(true) |> P$_1$ or  a$^2$(true) |> P$_2$ $\ldots$ or a$^n$(true)  |> P$_n$
or a$^1$(_) & a$^2$(_) & $\cdots$ & a$^n$(_) |> P$_0$
\end{lstlisting}
For each channel $\id{a}^i$ there are two forwarding channels
$\id{a}^i_{\kwd{true}}$ and $\id{a}^i_{\_}$.
As a consequence, after rewriting, the last reaction rule from the definition
above becomes:
\begin{lstlisting}{Join}
or (a$^1_\kwd{true}$(z$_1$) or a$^1_{\_}$(z$_1$)) & (a$^2_\kwd{true}$(z$_2$) or a$^2_{\_}$(z$_2$)) & $\cdots$ & (a$^n_\kwd{true}$(z$_n$) or a$^n_{\_}$(z$_n$)) |> P$_0$
\end{lstlisting}
And the expansion of \kwd{or} in join patterns finally yields $2^n$
reaction rules.

The extended \jocaml compiler indeed performs the expansion
of \kwd{or} in join patterns as sketched above, except for one point:
the guarded processes (\lst|P$_0$| in example) is not duplicated.
Instead, guarded processes are compiled into (lambda-code) closures
and duplication of guarded processes is performed by duplication of
pointers to those closures.

We will illustrate two successive refinements of the idea of sharing
guarded processes.  But before that, let us first examine how guarded
processes are compiled and triggered in the general case.
\begin{lstlisting}{Join}
def a(x) & b(y) |> $P$
or  a(x) & c(y) |> $Q$
\end{lstlisting}
The above join definition defines three channels organized in two
reaction rules. Target lambda-code can be sketched as follows:
\begin{lstlisting}[indent=1.8em,labelstep=1]{imppt}
let jdef =
    $\ldots\ldots$
  let g$_{\{a,b\}}$ = fun jdef ->                  $\label{lstGb}$ 
                let x = Join.get_queue jdef $i_{\id{a}}$ in $\label{lstGbx}$
                let y = Join.get_queue jdef $i_{\id{b}}$ in $\label{lstGby}$
                Join.unlock jdef;                        $\label{lstGbunlock}$
                Join.spawn (fun () -> $\CPLambda{P}$) in  $\label{lstGbspawn}$
  let g$_{\{a,c\}}$ = fun jdef ->                  $\label{lstGc}$ 
                let x = Join.get_queue jdef $i_{\id{a}}$ in $\label{lstGcx}$
                let y = Join.get_queue jdef $i_{\id{c}}$ in $\label{lstGcy}$
                Join.unlock jdef;                        $\label{lstGcunlock}$ 
                Join.spawn (fun () -> $\CPLambda{Q}$) in $\label{lstGcspawn}$
   $\ldots$
\end{lstlisting}
The presented lambda-code only describes the compilation of guarded
processes to closures $g_{\{a,b\}}$ and~$g_{\{a,c\}}$.  Those
\emph{guarded closures} are subparts of the complete compilation of
the join definition. They appear as local bindings in the more
complete definition \lst|jdef|, which is not shown.  We refer
to~\cite{LeFessantMarangetCompileJoin} for a full explanation about
how the \jocaml compiler deals with join definitions and guarded
processes.  Nevertheless, we give a brief description, based upon the
example.  Join definitions are compiled into vector-like structures,
and channels are pairs of a pointer to such a structure and of a
\emph{channel slot} (written $i_{\id{a}}$ etc. above).  Channel slots
are small integers.  Here, we assume $i_{\id{a}}$ to be~$0$,
$i_{\id{b}}$ to be~$1$, and $i_{\id{c}}$ to~be~$2$.  Based upon
channel slots, join patterns are compiled into bitsets. In this
example, we have $110$ for pattern ``\lst!a(x) & b(y)!''  and $101$
for ``\lst!a(x) & c(y)!''.  The join definition runtime structure
holds a list of pairs made of such a bitset and of a pointer to a
guarded closure (\lst|[($110$,g$_{\{a,b\}}$ ; ($101$,g$_{\{a,c\}}$)]|
in our example).  This \emph{join matching list} can be seen as the
result of reaction rules compilation.  The definition structure also
holds a mutex, an array of queues (indexed by channel slots), and an
internal bitset that describes the current status of queues.  In
response to message sending over a channel, specific code from the
\textsf{Join} library first locks the mutex, alters the internal
bitset, stores the message in the appropriate queue, and then attempt a
match.  In case a match is found, the corresponding closure
(\lst|g$_{\{a,b\}}$| or \lst|g$_{\{a,c\}}$| above) is called, with the
definition itself as an argument.

Notice that the closures \lst|g$_{\{a,b\}}$| or \lst|g$_{\{a,c\}}$|
have the responsibility to bind formal arguments \lst|x| and \lst|y|
to the
appropriate actual arguments, which are extracted from the appropriate
queues (lines~\ref{lstGbx}--\ref{lstGby}) and
\ref{lstGcx}--\ref{lstGcy}), and to release the mutex
(lines~\ref{lstGbunlock} and \ref{lstGcunlock}).
The guarded process is finally triggered by the means of
the primitive \lst|Join.spawn| that takes a closure as argument
(lines~\ref{lstGbspawn} and~\ref{lstGcspawn}) and creates a new thread
to run that closure.
Here,  $\CPLambda{P}$ and $\CPLambda{Q}$ represent the
compilation to lambda-code of $P$ and~$Q$ respectively.
It is to be noticed that formal parameters may occur free in
$P$ and~$Q$.

Now let us consider the compilation of join definitions with \kwd{or}
in their join patterns, such as this one:
\lst|def a(x) & (b(y) or c(y))|~$\triangleright~P$.
Target lambda-code  can be sketched as follows:
\begin{lstlisting}[indent=1.8em,labelstep=1]{bonga}
let jdef = 
   $\ldots\ldots$
  let p = fun jdef x y ->$\label{sharebegin}$
                Join.unlock jdef;
                Join.spawn (fun () -> $\CPLambda{P}$) in $\label{shareend}$
  let g$_{\{a,b\}}$ = fun jdef ->
                let x = Join.get_queue jdef $i_{\id{a}}$ in
                let y = Join.get_queue jdef $i_{\id{b}}$ in
                p jdef x y in $\label{call1}$
  let g$_{\{a,c\}}$ = fun jdef ->
                let x = Join.get_queue jdef $i_{\id{a}}$ in
                let y = Join.get_queue jdef $i_{\id{c}}$ in
                p jdef x y in $\label{call2}$
   $\ldots\ldots$
\end{lstlisting}
As a consequence of the expansion of the disjunctive pattern
``\lst"b(y) or c(y)"'', the join matching list is
\lst|[($110$,g$_{\{a,b\}}$ ; ($101$,g$_{\{a,c\}}$)]|, like in the
previous example. The two guarded closures $\id{g}_{\{a,b\}}$
and~$\id{g}_{\{a,c\}}$ are different, because the value bound to the
formal argument~$\id{y}$ has to be extracted either from the queue of
channel $\id{b}$ or from the queue of channel $\id{c}$, depending upon
the matched join pattern being ``\lst!a(x) & b(y)!'' or ``\lst!a(x) & c(y)!''.
However, the task of unlocking the mutex and of triggering
the process $P$ is common to both and is performed by a third
closure~\lst|p| (lines~\ref{sharebegin}--\ref{shareend}), which is
called by the two guarded closures $\id{g}_{\{a,b\}}$ and~
$\id{g}_{\{a,c\}}$ at lines \ref{call1} and~\ref{call2} respectively.
As a result, duplication of most of the guarded process code is
avoided and a reasonable amount of sharing is achieved.  One should
observe that the interface between the library code that performs join
matching and the guarded closures is preserved: guarded closures are
still functions that take a join structure as argument.

It is in fact possible for the compiler to completely share guarded
closures between reactions rules that originate from \kwd{or} pattern
expansion.  But then, guarded closure code must be abstracted further
with respect to the exact join pattern that is matched.  The idea of
dictionary can be used for this purpose. A \emph{dictionary} is
an array built by the compiler.  Dictionaries represent mappings from
formal parameters to channel slots and the compilation of a join
pattern now yields a pair of a bitset and of a dictionary.  More
significantly, disjunctive patterns are now compiled into a series of
such pairs.  For instance, the pattern ``\lst|a(x) & (b(y) or c(y))|''
is now compiled into the two pairs ``\lst!($110$,[|$0$ ; $1$|])!''
and ``\lst!($101$,[|$0$ ; $2$|])!'', where for instance the dictionary
component ``\lst![|$0$ ; $2$|]!'' expresses that the formal parameters
\lst|x| and~\lst|y| are to be bound to messages sent on channels
$\id{a}$ (at slot 0) and~$\id{c}$ (at slot 2) respectively.  The
compiler then generates guarded closures abstracted with respect to
dictionaries.
\begin{lstlisting}{Join}
let g$_{\{a,(b|c)\}}$ = fun jdef dict -> 
            let x = Join.get_queue jdef (field 0 dict) in
            let y = Join.get_queue jdef (field 1 dict) in
            Join.unlock jdef;
            Join.spawn (fun () -> $\CPLambda{P}$)
\end{lstlisting}
where ``\lst|field i dict|'' returns the $i$th element of the
dictionary ``\id{dict}''. The join matching list now becomes the
following list of triples:
\begin{lstlisting}{foo}
  [ ($110$,[|$0$ ; $1$|],g$_{\{a,(b|c)\}}$) ; ($101$,[|$0$ ; $2$|],g$_{\{a,(b|c)\}}$) ]
\end{lstlisting}
In case a join-pattern bitset is matched, the corresponding closure in
the triple is called, with the join definition structure and
additionally the dictionary in the triple as arguments.

Adding one dictionary component is the price we should pay to achieve
complete sharing of guarded closures. However, such a dictionary is
not necessary for reaction rules whose pattern is not disjunctive.  In
that case, the compiler can avoid the extra ``\lst|field i dict|''
calls and replace them by the appropriate channel slots, which are
known at compile time.  However, for the sake of keeping an uniform
structure of the join matching list, guarded closures should always
accept the extra ``\lst|dict|'' argument, even when not needed.  A
simple solution is to consider a dummy dictionary, to be passed to
such guarded closures that do not need a dictionary.

The current implementation of \jocaml does not use dictionaries.  We
are still lacking experience to be able to assert whether they are
worth the price or not.

\section{Related work}
\label{sec.relatedwork}

Applied join is ``impure'' in the sense of Abadi and Fournet's
applied $\pi$-calculus~\cite{AppliedPi}. We too extend an archetypal
name passing calculus with pragmatic constructs, in order to provide a
full semantics that handles realistic language features without
cumbersome encodings.  It is worth noticing that like
in~\cite{AppliedPi}, we distinguish between variables and names (only
variables of channel type are treated as names), a distinction that is
seldom made in pure calculi. Since we aim to prove a program
transformation correct, we define the equivalence on open terms, those
that contain free variables. Abadi and Fournet are able to require
their terms to have no free variables, since their goal is to prove
properties of program execution, namely the correctness of security
protocols.

Our compilation scheme presented in Section~\ref{sec.trans} can be
seen as the combination of two basic steps: refining channels and
forwarding by dispatcher. The desired property of the forwarding
behavior (Lemma~\ref{lemma.joinpi.dispatcher}) constitutes the core of
the correctness proof of the compilation scheme, which essentially
stems from pattern matching theory. There are other work that perform
the formal treatments of forwarders, for
instance~\cite{Merro98asynchrony,Gardner2003lf}, but in different
contexts. Our forwarder demultiplexes messages into separate channels
according to the pattern of the messages,
while~\cite{Merro98asynchrony,Gardner2003lf} use plain
channel-to-channel linear forwarders to achieve the locality property,
\ie reception on a given channel takes place on an unique site.
It is to be noticed that the equivalence proof of~\cite{Gardner2003lf}
is with respect to ordinary barbed congruence
and by the means of a labelled transition system.
Yet another example is the correctness proof of the compilation of
join patterns to smooth orchestrators in~\cite{Orchestrators}.
The compilation of~\cite{Orchestrators} is less involved than ours
since it basically amounts to inserting forwarders.

We established the correctness of our compilation scheme by showing
the programs before and after compilation to be behavioral congruent. It is
usual practice in the literature to prove correctness of program
transformations by showing semantics preservation.
(\cite{DaveCompVeriSurvey} is a survey).
Here, variations are numerous:
they consist in different connections between source and
target formalism (two independent languages, or with the target being
a sub-set of the source), different semantics (denotational vs.
operational), different equivalence relations (observational
equivalence, refinement relation, simulation, etc.), and different
settings (sequential, concurrent, parallel, object-oriented, etc.),
Consequently, proof techniques also differ.
For example, recent work of Blazy~\etal~\cite{Blazy-Dargaye-Leroy-06}
reports the formal verification of a C compiler front-end in the Coq
proof assistant. It handles two independent source and target
languages, both with big-step operational semantics. The major
difficulty of the correctness proof resides in relating the different
memory states and evaluation environments of the two languages. A
simulation relation is demonstrated from target code to source code by
induction on evaluation derivation and case study over the last
applied evaluation rule.
Closer to our work, \cite{EmirMaOdersky2007} shows the correctness
of an optimizing translation that compiles away pattern matching in
\scala. Proof techniques analogous with ours are applied, \ie they
also tackle contexts explicitly by proving congruence and define
observational equivalence on open terms based on the one between
closed terms and closing up by substitutions. Moreover, specific to
its extractor-base pattern matching, extractors are required to always
terminate without exception in order to achieve the correctness.

We now review some programming languages that support concurrency and
examine how our work can be related to those. Languages whose model
for concurrency directly stems from the join calculus should benefit
from our work. More precisely, if a language already offers \`{a} la
ML pattern matching and join definitions, then its authors can
implement our ideas in their framework, and their implementation
effort would be small. An early example of a language based upon the
join calculus is \funnel~\cite{funnel}. \funnel later evolved into
\scala~\cite{Odersky:scala}, where \`{a} la ML pattern matching is
supported and join style concurrency is provided in terms of a
library~\cite{JoinScala}. Another similar work is~\cite{Singh06},
which introduces join style concurrency in \haskell. We believe that
extending the two settings above
with algebraic patterns as formal arguments
can be made by direct application of our techniques.
Smooth orchestrators~\cite{Orchestrators} differ from join definitions
in rather subtle ways: an orchestrator is syntactically similar to a
join definition and can be seen as defining competing reaction rules;
however, (1)~once a reaction rule of an orchestrator is selected and
continuation fired, the whole orchestrator (together with other
non-selected competing rules) gets expired and discarded; and~(2) the
definitions of channels and of orchestrators that synchronize them are
separated.
Point~(2) above is quite subtle: one can orchestrate receptions
on channels whose definitions are unknown, provided
all the orchestrated channels are defined on the same site.
Nevertheless,
orchestrators are controlled by finite automata that extends the ones
of~\cite{LeFessantMarangetCompileJoin} for join~definitions.  Thus,
the adaptation of our techniques to orchestrators looks feasible.

In addition, there is a sustained interest in integrating join
calculus into object-oriented languages~: polyphonic~\csharp and its
successor \comega~\cite{Cw} for \csharp; and \joinjava~\cite{JoinJava}
for \java. Unfortunately, the issue here is the lack of pattern
matching, which neither \csharp nor \java offers.  A detailed
discussion on the introduction of \`{a} la ML pattern matching in
object-oriented languages would be out of scope.  Briefly, proposed
solutions are either by the means of preprocessing~\cite{TOM}, or by
tighter language integration~\cite{Odersky:scala,OOMatch}. As our
compilation scheme requires precise information on pattern semantics
(\eg to decide the precision relation~$\preceq$), we think
that solutions of the second kind would facilitate the extension of
the introduced pattern matching to join patterns.

\erlang~\cite{Erlang} features both pattern matching and concurrency.
However, concurrency in Erlang is based upon the actor
model~\cite{actor73,actor86}. In this model, messages are sent to
\emph{actors} and actors manage a queue of messages.
Moreover, the
reception behavior of an actor can be specified by
the  \texttt{receive}~\textit{m} construct.
This construct is similar to ML pattern matching
\texttt{match}~\textit{v}~\texttt{with}~\textit{m}, except for the
value matched~$v$, which is left implicit.
The semantics of \texttt{receive}~\textit{m}
can be described as follows: attempt a match in the actor's queue,
scanning it from the oldest to the most
recent message, stopping when a match is found.
This simple combination of message passing and pattern
matching proves convenient, as witnessed by the success of Erlang.
However, \erlang in general misses a simple and efficient handling of
synchronization between actors as join patterns offer. Lacking
necessary knowledge of \erlang internals, it is difficult for us to
assess whether the selection of messages from actors queues can
benefit from our techniques or not. In any case, difference in
semantics is outstanding and we conjecture that an adaption of our
technique would not be immediate. In particular, the existence of one
message queue per receiving agent is central to \erlang model, while a
join definition naturally handles several message queues.

Finally, we discuss the transplantation of our compilation scheme to a
language whose semantics for concurrency is based upon the original
$\pi$-calculus of~\cite{MPW92}, like for instance
\textrm{Pict}~\cite{Pict}, or \textrm{PiDuce}~\cite{PiDuce:2005}
without orchestrators.  Such a task is apparently impossible.  Namely,
on the one hand, we propose a \emph{compilation} scheme, and we thus
need to isolate all the instances of reception on a given channel from
program source~; while, on the other hand, the $\pi$-calculus features
\emph{unrestricted input capability}.
More precisely, in the  $\pi$-calculus,
any process that knows of some
channel~$x$ can input on it.  As a channel name~$x$ can be passed via
messages, reception on~$x$ may occur anywhere.  The join~calculus
originates from a radical solution to the distributed implementation
issue: channels and reception behaviors are defined by a synthetic
construct, and input on channels cannot occur anywhere else.  However,
there are other solutions that retain the $\pi$-calculus as a basis
while restricting input capability, such as the localized
$\pi$-calculus~\cite{Merro98asynchrony}.
Moreover, the located channels of Nomadic Pict~\cite{nomadic}
allows to lift such solutions to a distributed setting.
Given such frameworks, we
shall assume that all receptors on a given channel are known
statically.  Then, we can extend the input construct $x(y).P$ as
$x(\pt).P$, where $\pt$ is pattern, and expect to be able to translate
this extended language into ordinary $\pi$-calculus.  In that process,
we see at least one additional complication.  Let $\pt_1$ and~$\pt_2$
be two patterns that are compatible (\ie that have instances in
common), and let us consider the following program, an analog of the
simple examples of Section~\ref{sec.trans-idea}.
$$
x(\pt_1).P_1 \mid x(\pt_2).P_2
$$
The above process significantly differs from a join definition, since
a successful input does not discard the other input.
A tentative translation in the spirit of ours would
be the parallel composition of a dispatcher:
$$
!~x(z).\texttt{match}\ z\ \texttt{with}\
\lub{\pt_1}{\pt_2} \rightarrow x_{\lub{\pt_1}{\pt_2}}(z)\ 
\texttt{|}\ \pt_1 \rightarrow x_{\pt_1}(z)\ 
\texttt{|}\ \pt_2 \rightarrow x_{\pt_2}(z)
$$
and of the following process:
$$
(x_{\lub{\pt_1}{\pt_2}}(z).Q_1 + x_{\pt_1}(z).Q_1) \mid
(x_{\lub{\pt_1}{\pt_2}}(z).Q_2 + x_{\pt_2}(z).Q_2)
$$
Where $Q_i$ is $\texttt{match}~z~\texttt{with}~\pt_i \rightarrow P_i$,
and ``$+$'' is internal choice that we use here to express input-guarded choice.
Thus, we need input-guarded choice.
This is a noticeable complication, even though
input-guarded choice can be expressed
in the $\pi$-calculus without choice~\cite{NestmannPierce00}.
Another concern is the usage of the replication operator~``$!$''
in the dispatcher.
Clearly, the adaptation of our technique to a $\pi$-calculus setting
is not immediate.

\section{Conclusion and future work}
\label{sec.conclusion}

This paper is part of our effort to develop a practical concurrent
programming language with firm semantical foundations.
In our opinion, a programming language is more
than an accumulation of features. That is, features interact sometimes
in unexpected ways, especially when intimately entwined.  Here, we
have studied the interaction between pattern matching and concurrency.
The framework we have used was the applied join calculus --- an
extension of the join calculus with algebraic data types. Applied join
inherits its capabilities of communication and concurrency from join
and supports value passing.  More significantly, it allows algebraic
pattern matching in both formal arguments of channel definitions and
guarded processes.  Compared with join, applied join provides a more
convenient (or ``pragmatic''), precise and realistic language model to
programmers.  From that perspective, pattern matching and join
calculus appear to live well together, with mutual benefits. The
result of this work reinforces our interest in using \`{a} la ML
pattern matching as a general purpose programming paradigm, and join
calculus as the basic paradigm for concurrency.

Exploiting the fact that \jocaml already had an efficient
implementation for both ML pattern matching and join primitives, we
have designed the implementation of applied join as defining a
practical compilation scheme that transforms extended join definitions
into ordinary ones plus ML pattern matching. We have solved the
non-determinism problem during the design of this compilation scheme.
Moreover, we have actually integrated it into the \jocaml system with
several optimizations.
It is worth observing that a direct
implementation of extended join-pattern matching at the runtime level
would significantly complicate the management of message queues, which
would then need to be scanned in search of matching messages before
consuming them.
As we remarked, our compilation technique may yield
code of exponential size.
However, we expect such blowup not to occur
in practice, an expectation which is apparently confirmed by
our preliminary experiments in the \jocaml system.
Should this prove wrong in the future, we could
face the issue in two manners~: either complicate the runtime system
as sketched above, or design a direct implementation of \kwd{or} in
join patterns.

A theory of process equivalence has also been developed in applied
join in order to assess the correctness of our compilation scheme. In
archetypal name passing calculi, where every free variable is of
channel type, it is sufficient to only consider terms closed in our
sense, \ie terms without free variables of non-channel type, when
defining equivalence relations. By contrast, applied join supports
real values and its static transformations should apply to open
processes with free variables of non-channel type.  To tackle this
problem, we have first defined a weak barbed congruence to express the
equivalence of two closed processes, then we have lifted the
equivalence relation to open processes by closing up by all
substitutions. The resulting relation is called ``open equivalence''.
We have demonstrated it is also a full congruence and have proved our
compilation scheme correct by showing that the processes before and
after transformation are open equivalent.  The proof technique we have
used, which can be summarized as ``full abstraction'', stems from
pattern matching theory and the fact that inserting an internal
forwarding step in communications does not change process behavior.

In previous work, we have designed an object-oriented extension of the
join
calculus~\cite{FournetLaneve03,MaMaranget2003type,MaMaranget2005hideTR},
which appeared to be more difficult. The difficulties reside in the
refinement of the synchronization behavior of objects by using the
inheritance paradigm. We solved the problem by designing a delicate
way of rewriting join patterns at the class level.  However, the
introduction of algebraic patterns in join patterns impacts this
class-rewriting mechanism. The interaction is not immediately clear.
Up to now, we are aware of no object-oriented language where the
formal arguments of methods can be patterns. We thus plan to
investigate such a combination of pattern matching and inheritance,
both at the calculus and language level.

Another interesting future work would be to extend our framework with
more sophisticated patterns for XML data. As a matter of fact, the
authors of \scala have already extended the notion of pattern matching
to the processing of XML data with the help of regular expression
patterns (a similar system is \textrm{PiDuce}~\cite{PiDuce:2005}).  Their
extension makes \scala suitable for developing web service
applications. Our model of pattern matching in join calculus works
with general algebraic data types. At the moment, we do not see any
particular barrier that prevent our model from also working with XML
trees.

\section*{Acknowledgement}
The authors wish to thank James Leifer and Jean-Jacques L\'{e}vy for
fruitful discussions and comments.
We also thank the anonymous referees for their suggestions.

\bibliographystyle{plain}
\bibliography{ptjoin}
\end{document}